\documentclass[a4paper,11pt]{article}
\usepackage{graphicx}
\usepackage{jheppub}
\usepackage[above]{placeins}
\usepackage[T1]{fontenc} 
\title{\boldmath Decoherence and disentanglement of qubits  detecting  scalar fields in an expanded universe}
\author[a]{Yujie Li}
\author[a]{Yue Dai}
\author[a,b,1]{Yu Shi\note{Corresponding author.}}
\affiliation[a]{Center for Field Theory and Particle Physics, Department of Physics  \&  State Key Laboratory of Surface Physics,   Fudan University,\\ Shanghai 200433, China}
\affiliation[b]{Collaborative Innovation Center of Advanced Microstructures, Fudan University, \\Shanghai 200433, China}
\emailAdd{liyujie800@163.com}
\emailAdd{dy1983@gmail.com}
\emailAdd{yushi@fudan.edu.cn}

\abstract{ We consider Unruh-Wald qubit detector model adopted for the far future region of an exactly solvable 1+1 dimensional scalar field theory in a Robertson-Walker expanding spacetime. It is shown that the expansion of the universe in its history  enhances the decoherence of the qubit coupled with a scalar field. Moreover, we consider two  entangled qubits, each locally coupled a scalar field. The expansion of the universe in its history  degrades the entanglement between the qubits, and can lead to entanglement sudden death if the initial entanglement is small enough.  The details depend on the parameters characterizing the expansion of the universe. This work, albeit on a toy model, suggests that the history of the universe might be probed through the coherent and entanglement behavior of future detectors of quantum fields.}

\begin{document}
\maketitle
\flushbottom

\section{Introduction}

Recently, the concepts developed in quantum foundations and quantum information theory have been exploited to understand quantum effects of spacetime, including  the quantum effects of the expansion of the  universe~\cite{em}. Concepts such as quantum entanglement can shed new light on the topic of the particle creation in an expanding universe~\cite{parker1,parker2}.  Investigations have been made on the entanglement generated  between different field modes by the expansion of a model universe, and scalar, Dirac and some other fields have been studied~\cite{ball,fuentes,moradi,mohammadzadeh,wang}.  It was also shown that the entanglement in the field  can be swapped to   detectors~\cite{steeg}. The response of a detector switched on since the early universe were  also studied~\cite{garay}. Basic issues  concerning the entanglement created in a time-dependent spacetime were carefully examined and clarified~\cite{lin}.

In this paper, we investigate the cosmological effect on the coherence and entanglement of detectors, rather than field modes. Specifically, we consider an exactly solvable model of scalar field in an expanding universe~\cite{davies}, which is a common model of field theory used  in the present subject.

The field theory model and the qubit detector model are introduced in  Secs.~\ref{field} and \ref{detector}, with the single mode approximation justified. The decoherence of a single qubit is discussed in Sec.~\ref{single}, by studying the dependence of its purity on the two parameters characterizing the expansion of this model universe. Then in Sec.~\ref{two}, we move on to two initially entangled qubits,  studying the mutual information, which is a quantifying measure of the total correlation including both classical correlation and quantum entanglement, and the concurrence, which is a quantifying measure of entanglement. Afterwards, in Sec.~\ref{teleportation}, we consider quantum teleportation  under the coupling with the fields in the expanded spacetime, and calculate its fidelity. In Sec.~\ref{explain}, we explain the common features in the dependence of different quantities on the two cosmic parameters.   Finally, we make summary and discussions in Sec.~\ref{summary}.

\section{Scalar field in a model of expanding universe \label{field} }

Consider a $1+1$ dimensional   Robertson-Walker metric, the line element being  $ds^{2}=dt^{2}-a^{2}\left(t\right)dx^{2}$,
where $a\left(t\right)$ is the scale factor. By using the conformal time $\eta$ defined as  $d\eta=\dfrac{dt}{a\left(t\right)}$,
the line element is rewritten as
\begin{equation}
ds^{2}=R^{2}\left(\eta\right)\left(d\eta^{2}-dx^{2}\right). \end{equation}
Suppose the the conformal scale factor is~\cite{ball}
\begin{equation}
 R^{2}\left(\eta\right)=1+\varepsilon[1+\tanh\left(\sigma\eta\right)], \label{metric}
\end{equation}
with the parameters $\varepsilon$ and $\sigma$  characterizing the volume and the rapidity of the  expansion of the universe, respectively.  It can be seen that the spacetime is flat in the distant past and in the far future, that is, $ds^{2}= d\eta^{2}-dx^{2}$ as $\eta \rightarrow -\infty$, while $ds^{2}= \left(1+2\varepsilon\right)\left(d\eta^{2}-dx^{2}\right)$ as  $\eta \rightarrow +\infty$. Consequently, the timelike Killing vector and thus the particle content of the field are  well defined in these two limits.

In this metric, consider a real scalar field $\Phi\left(x,\eta\right)$, which  satisfies the Klein-Gordon  equation
\begin{equation}\left( \Box  +m^{2}\right)\Phi=0, \end{equation}
with $\Box  \Phi \equiv \partial_\mu\left(\sqrt{-g}g^{\mu\nu}\partial_\nu \Phi\right)/\sqrt{-g}$. Corresponding to the limits of $\eta \rightarrow \pm \infty$, there exist a set of  basis solutions $u^{{\mathrm{in}}}_\mathbf{k}$  in the distant past or ``in'' region, and  a set of  basis solutions  $u^{{\mathrm{out}}}_\mathbf{k}$   in the far  future  or ``out'' region, and they are related as~\cite{bernard,davies}
\begin{equation}
u^{{\mathrm{in}}}_\mathbf{k}\left(x,\eta\right)
  =\alpha_\mathbf{k}u^{{\mathrm{out}}}_\mathbf{k}\left(x,\eta\right)+ \beta_\mathbf{k}u^{{\mathrm{out}}*}_{-\mathbf{k}}\left(x,\eta\right),\end{equation}
where
  \begin{equation}\alpha_\mathbf{k}\equiv \left(\dfrac{\omega_{{\mathrm{out}}}}{\omega_{{\mathrm{in}}}}\right)^{1/2}\dfrac{\Gamma\left(1-\dfrac{i\omega_{{\mathrm{in}}}}{\sigma}\right)\Gamma\left( -\dfrac{i\omega_{{\mathrm{out}}}}{\sigma}\right)}{\Gamma\left( -\dfrac{i\omega_{+}}{\sigma}\right)\Gamma\left( 1-\dfrac{i\omega_{+}}{\sigma}\right)},\end{equation}
   \begin{equation}\beta_\mathbf{k}\equiv \left(\dfrac{\omega_{{\mathrm{out}}}}{\omega_{{\mathrm{in}}}}\right)^{1/2}\dfrac{\Gamma\left( 1-\dfrac{i\omega_{{\mathrm{in}}}}{\sigma}\right)\Gamma\left( \dfrac{i\omega_{{\mathrm{out}}}}{\sigma}\right)}{\Gamma\left( \dfrac{i\omega_{-}}{\sigma}\right)\Gamma\left( 1+\dfrac{i\omega_{-}}{\sigma}\right)},\end{equation}
with
$\omega_{{\mathrm{in}}}=[k^{2}+m^{2}]^{1/2},$
   $\omega_{{\mathrm{out}}}=[k^{2}+m^{2}\left(1+2\varepsilon\right)]^{1/2}$,
$\omega_{\pm} \equiv
\dfrac{1}{2}\left(\omega_{{\mathrm{out}}}\pm\omega_{{\mathrm{in}}}\right)$, $\Gamma$ being the gamma function, $k\equiv |\mathbf{k}|$. It can be obtained that
$|\alpha_\mathbf{k}|^{2}=
   \dfrac{\sinh^{2}\left(\pi\omega_{+}/\sigma\right)}{\sinh\left(\pi\omega_{{\mathrm{in}}}/ \sigma\right)
   \sinh\left(\pi\omega_{{\mathrm{out}}}/\sigma\right)}$, $|\beta_\mathbf{k}|^{2}=
   \dfrac{\sinh^{2}\left(\pi\omega_{-}/\sigma\right)}{\sinh\left(\pi\omega_{{\mathrm{in}}}/\sigma\right)
   \sinh\left(\pi\omega_{{\mathrm{out}}}/\sigma\right)}$,  satisfying  $|\alpha_\mathbf{k}|^{2}-|\beta_\mathbf{k}|^{2}=1$.  For convenience, we define
 \begin{equation}
 \gamma_k  = { \left| {\frac{{{\beta _{\mathbf{k}}}}}{{{\alpha _{\mathbf{k}}}}}} \right|^2} =   \frac{{{{\sinh }^2}\left( {\pi {{{\omega _ - }} \mathord{ /
 {\vphantom {{{\omega _ - }} \sigma }}\kern-\nulldelimiterspace} \sigma }} \right)}}{{{{\sinh }^2} \left( {\pi {{{\omega _ + }} \mathord{ /
 {\vphantom {{{\omega _ + }} \sigma }}\kern-\nulldelimiterspace} \sigma }} \right)}}, \end{equation}
which measures the degree of mixing between the ``in'' modes $\mathbf{k}$ and $-\mathbf{k}$. It also measures the average number of particles created at ``out'' mode $\mathbf{k}$, which equals $|\beta_\mathbf{k}|^2$~\cite{parker2}. 
\begin{equation} 
\gamma_k = \frac{|\beta_{\mathbf{k}}|^2}{1+|\beta_{\mathbf{k}}|^2},
\end{equation}
\begin{equation} 
|\beta_\mathbf{k}|^2 = \frac{1}{\gamma_k^{-1}-1}.
\end{equation}
Hence $\gamma_k \rightarrow 0$ means that the average number of the particles created at the ``out'' mode $\mathbf{k}$ is vanishing, $\gamma_k \rightarrow 1$ means that  the average number of the particles created at the ``out'' mode $\mathbf{k}$ approaches infinity.

The annihilation and creation operators satisfy
\begin{equation}
\hat{a}^{{\mathrm{in}}}_\mathbf{k}=
\alpha^{*}_\mathbf{k}\hat{a}^{{\mathrm{out}}}_\mathbf{k}-
\beta^{*}_\mathbf{k}\hat{a}^{{\mathrm{out}}\dagger}_{-\mathbf{k}},\end{equation}
\begin{equation}\hat{a}^{{\mathrm{in}}\dagger}_\mathbf{k}=
\alpha_\mathbf{k}\hat{a}^{{\mathrm{out}}\dagger}_\mathbf{k} - \beta_\mathbf{k}\hat{a}^{{\mathrm{out}}}_{-\mathbf{k}}.
\label{ain} \end{equation}

As the Bogoliubov transformation only mixes the ``out'' modes $\mathbf{k}$ and $-\mathbf{k}$  in the  ``in'' mode   $\mathbf{k}$, the  ``in'' vacuum  in the sector for the (unordered) pair  $\mathbf{k}$ and $-\mathbf{k}$   is thus
\begin{equation}
|0\rangle_\mathbf{k}^{\mathrm{in}}|0\rangle_{-\mathbf{k}}^{\mathrm{in}}
= \sum_n
A_{n,\mathbf{k}} |n\rangle_\mathbf{k}^{\rm{out}} |n\rangle_{- \mathbf{k}}^{\rm{out}},
\label{vac}
\end{equation}
where $n$ denotes the particle number,  $A_{n,\mathbf{k}} = { \left( {\frac{{\beta _{\mathbf{k}}^*}}{{\alpha _{\mathbf{k}}^*}}} \right)^n}\sqrt {1 - \gamma_k}$. After creation, each pair of modes with opposite  momenta are separated on  cosmological scale~\cite{parker2}.    A local detector at the far future accesses only one of each paired modes, say $\mathbf{k}$, hence feels a mixed state, with the other mode  $-\mathbf{k}$ traced out. In (\ref{vac}),  the reduced density matrix of mode $\mathbf{k}$ is
 \begin{equation}
 \rho_\mathbf{k}^{\rm{out}}  = \mathrm{Tr}_{ -\mathbf{k} }  [| 0 \rangle_{\mathbf{k}}^{\mathrm{in}} | 0 \rangle_{-\mathbf{k}}^{\mathrm{in}} {_{-\mathbf{k}}^\mathrm{in}} \langle 0  | {_{\mathbf{k}}^\mathrm{in}} \langle 0  | ]
  =  \left( 1 - \gamma_k\right) \sum\limits_n  \gamma_k^n  |n \rangle_{\mathbf{k}} \langle n |,
 \end{equation}
 where   $|n \rangle_{\mathbf{k}}\langle n |$ on the rightmost is a shorthand for  $|n \rangle_{\mathbf{k}}^{\mathrm{out}}  {_{\mathbf{k}}^{\mathrm{out}}} \langle n |$. From now on, the superscript ``out'' is dropped without causing confusion. The density matrix of all the modes accessible to the local detector is $\mathop \prod \limits_{\mathbf{k}} {\rho _{\mathbf{k}}^{\rm{out}}}$,
where  the direct product is only over those accessible modes.

\section{Coupling between the detector qubit and the scalar field \label{detector} }

Suppose at the far future of the expanded universe described by (\ref{metric}), a  detector couples locally with a scalar field. For simplicity, we adopt for the present purpose the Unruh-Wald qubit detector  model, which is originally a detector model for Unruh effect~\cite{unruhwald}. The Hamiltonian is
$H_{\Phi}+H_q+H_{I}$, where
$H_{\Phi}$ is the Klein-Gordon Hamiltonian for the scalar field. $H_q$ is the Hamiltonian of the qubit detector,   given by
\begin{equation}
H_q=\Omega Q^\dagger Q, \label{qubit}
\end{equation}
where $Q^\dagger$ and  $Q$ are creation and annihilation operators acting on two basis states $|0\rangle$  and $|1\rangle$ of the qubit as  $Q | 0 \rangle = Q^\dagger  | 1 \rangle = 0,$ $Q^\dagger  | 0 \rangle =  | 1 \rangle$,
$Q | 1 \rangle = | 0\rangle$. $\Omega$ is  the energy level difference between $|1\rangle$ and $|0\rangle$. The interaction $H_{I}$ is
\begin{equation}
H_{I}\left(t\right) = \epsilon \left( t \right) \int_{\Sigma}  {\Phi  \left(  \mathbf{x},t \right) [ {\psi  \left( \mathbf{ x}  \right)Q  + {\psi^*} \left( \mathbf{x} \right){Q^\dag }}  ]\sqrt { - g} {d}x}.
\end{equation}
where $\mathbf{x}$ and $t$ are proper  coordinates  of the qubit,  the integral is over the spacelike Cauchy surface $\Sigma$ at given time $t$, $\epsilon\left( t \right)$ is the coupling constant with a finite duration of qubit-field interaction, $\psi \left(\mathbf{ x} \right)$ is a smooth function nonvanishing within a small volume around the qubit. In the interaction picture, the unitary transformation induced by the  Hamiltonian can be written as~\cite{unruhwald}
\begin{equation}
U \approx 1-i \int \Phi\left(\mathbf{x},t'\right)\epsilon\left(t'\right) [Q e^{-i\Omega t'} \psi\left(\mathbf{x}\right) + Q^\dagger e^{i\Omega t'} \psi^*\left(\mathbf{x}\right)]\sqrt{-g'} dx dt',
\end{equation}
which, because of resonant effect, can be simplified as
\begin{equation}
U \approx 1+iQ a^\dagger \left(\Gamma^*\right)  -  i  Q^\dagger a\left(\Gamma^*\right),
\end{equation}
where $a\left(\Gamma^*\right)$  and $a^\dagger\left(\Gamma^*\right)$ are the  annihilation and  the creation operators of the mode $\Gamma_q^*$, with
\begin{equation}
\Gamma \left(x\right) \equiv -2i \int [G_R\left(x;x'\right)-G_A\left(x;x'\right)]\epsilon\left(t'\right)e^{i\Omega t'}\psi^*\left(\mathbf{x}'\right)\sqrt{-g'}
d^2x',
\end{equation}
$G_{R}$ and $G_{A}$ being the retarded and advanced Green functions of the field $\Phi$, respectively. We stay in the interaction picture, in which the results we shall be interested in are the same as those in the Schr\"{o}dinger picture.

For each mode $\chi_\mathbf{k}$, the action of $a\left(\Gamma^*\right)$ and  $a^\dagger \left(\Gamma^*\right)$ is
\begin{eqnarray}
a\left(\Gamma^*\right)|n\rangle_\mathbf{k} &=& \sqrt{n} \mu_\mathbf{k} |n-1\rangle_\mathbf{k},\\
a^\dagger \left(\Gamma^*\right)|n\rangle_\mathbf{k} &=& \sqrt{n+1} \mu_\mathbf{k}^* |n+1\rangle_\mathbf{k},
\end{eqnarray}
where  $\mu_\mathbf{k}\equiv \langle \Gamma_q^*,\chi_\mathbf{k}\rangle =  \int \epsilon_q\left(t\right)  e^{i\Omega_q t}\psi_q^*\left(\mathbf{x}\right) \chi\left(t,\mathbf{x}\right)\sqrt{-g} d^2 x$ is the inner product~\cite{unruhwald}.

As a single mode approximation, we may ignore the coupling between the detector qubit and all the field modes except the mode $\mathbf{k}_0$ with energy equal to $\Omega$. $-\mathbf{k}_0$ is out of access because of separation on cosmological scale.

Therefore, the effect of the interaction between the detector qubit and the field mode  $\mathbf{k}_0$  can be simplified as
\begin{equation}
\begin{array}{l}
 | n \rangle  \otimes  | 0 \rangle  \to  | n \rangle  \otimes  | 0 \rangle  - i\sqrt n \mu | {n - 1} \rangle  \otimes  | 1 \rangle, \\
 | n \rangle  \otimes  | 1 \rangle  \to  | n \rangle  \otimes  | 1 \rangle  + i\sqrt {n + 1} {\mu ^*} | {n + 1} \rangle  \otimes  | 0 \rangle,
\end{array}   \label{n}
\end{equation}
where we omit the mode indice $\mathbf{k}_0$ of the Fock state $|n\rangle$ and the inner product
\begin{equation}
\mu\equiv \langle \Gamma_q^*,\chi_{\mathbf{k}_0}\rangle.  \label{mu0}
\end{equation}

For a particle created by the expansion of the universe, the lower bound of the energy is $m\sqrt {1 + 2\varepsilon } $, as ${\omega_\mathrm{out}} = \sqrt {{k^2} + {m^2}\left( {1 + 2\varepsilon } \right)} \ge m\sqrt {1 + 2\varepsilon } $.   In order that  the qubit is affected by the field modes, there must be
\begin{equation}
\Omega  \ge m\sqrt {1 + 2\varepsilon }, \label{omega}
\end{equation}
that is,
\begin{equation}
\varepsilon \leq \varepsilon_{max} = \frac{1}{2}\left(\frac{\Omega^2}{m^2}-1\right).
\end{equation}
For example,   for $\Omega=2$ and $m=0.01$, in order for the field to be coupled with the qubit, the maximal volume of the expansion of the universe is ${\varepsilon _{\max }} = 19999.5$.

\section{Decoherence of a single qubit  \label{single} }

We now consider a single qubit detector. Its initial state is
\begin{equation}
 | \psi  \rangle  = \alpha  | 0 \rangle  + \beta  | 1 \rangle, \label{initial}
\end{equation}
with ${ | \alpha  |^2} + { | \beta  |^2} = 1$.

In the case that there has not been  expansion of the universe, the state of the field remains as the vacuum $|0\rangle$ when its interaction with the qubit switched on. In order to be coupled with the field, the qubit must satisfy $\Omega \geq m$. Then  as a special case of (\ref{n}), we have
\begin{equation}
 \begin{array}{l}
 | 0 \rangle  \otimes  | 0 \rangle  \to  | 0 \rangle  \otimes  | 0 \rangle, \\
 | 0 \rangle  \otimes  | 1 \rangle  \to  | 0 \rangle  \otimes  | 1 \rangle  + i {\mu ^*} |  1 \rangle  \otimes  | 0 \rangle.
\end{array}   \label{n0}  \end{equation}

Therefore the state of the field mode  $\mathbf{k}_0$  and the qubit evolves as
\begin{equation}
| 0 \rangle   \otimes  \left(\alpha  | 0 \rangle  + \beta  | 1 \rangle\right)  \rightarrow | 0 \rangle \otimes \left(\alpha  | 0 \rangle  + \beta  | 1 \rangle\right)  + i\beta {\mu ^*} |  1 \rangle  \otimes  | 0 \rangle.
\end{equation}
If $\beta \neq 0$, the qubit becomes entangled with the field. The reduced density matrix of the qubit is
\begin{equation}
\rho  = \frac{1}{1+|\beta|^2|\mu|^2}  \left( \begin{array}{cc}
|\alpha|^2+|\beta|^2|\mu|^2 & \alpha \beta^* \\
\alpha^* \beta & |\beta|^2 \end{array}  \right).   \label{rho1} \end{equation}

Its mixedness can be characterized by its  purity
\begin{equation}
\mathrm{Tr}(\rho^2)= \frac{(|\alpha|^2+|\beta|^2|\mu|^2 )^2+2|\alpha|^2|\beta|^2+|\beta|^4}{(1+|\beta|^2|\mu|^2)^2},
\end{equation}
which reduces to unity when $\beta=0$, then the state of the field and the qubit remains as the initial state $| 0 \rangle   \otimes  | 0 \rangle $.

We now turn to the case that there has been expansion of the universe. The particles generated by the expansion of the universe become the environment of the detector, and causes the decoherence of the detector. After the expansion, the qubit starts with the initial state (\ref{initial}). Hence  the state of the qubit and the field  mode  $\mathbf{k}_0$  starts as the separable mixed state
     \begin{equation}
{\rho _{\mathbf{k}_0}} \otimes  | \psi  \rangle  \langle \psi  |
=  \left( {1 - \gamma } \right)\sum\limits_n^{} {{\gamma ^n} [ { | n \rangle    \left( {\alpha  | 0 \rangle  + \beta  | 1 \rangle } \right)} ]    [ { \langle n |   \left( {{\alpha ^*} \langle 0 | + {\beta ^*} \langle 1 |} \right)} ]}.
 \end{equation}

According to (\ref{n}), after the interaction between the qubit and scalar field,
\begin{equation}
 | n \rangle    \left( \alpha  | 0 \rangle  + \beta  | 1 \rangle  \right) \rightarrow
 |\phi\rangle \equiv \frac{1}{{\sqrt {{Q_n}} }} [  | n \rangle  \left( \alpha  | 0 \rangle  + \beta  | 1 \rangle  \right) - i\alpha \sqrt n \mu  | {n - 1} \rangle  | 1 \rangle  + i\beta \sqrt {n + 1} {\mu ^*} | {n + 1} \rangle | 0 \rangle ],
 \end{equation}
where ${Q_n} = 1 + { | \alpha  |^2}n{ | \mu  |^2} + { | \beta  |^2} \left( {n + 1} \right){ | \mu  |^2}$ is the normalization factor. Hence the state of the field mode $\mathbf{k}_0$ and the qubit becomes
 \begin{equation}
\left(1 - \gamma\right) \sum_n  \gamma^n |\phi\rangle \langle \phi|.
 \end{equation}

By tracing out the field mode, one obtains the density matrix of the final state of the qubit,
\begin{equation}
\rho = \left( 1 - \gamma  \right) \sum_n \frac{\gamma^n}{Q_n} [ \left(\alpha | 0 \rangle  + \beta | 1 \rangle \right) \left( \alpha^*\langle 0 | + \beta^*\langle 1 |  \right)     +   | \beta|^2 \left( {n + 1} \right)| \mu  |^2 | 0 \rangle \langle 0 | + | \alpha  |^2 n | \mu  |^2 | 1 \rangle \langle 1 |  ],
\end{equation}
which, in the basis $\{|0\rangle, |1\rangle \}$, can be written as
\begin{equation}
\rho = \left( {\begin{array}{*{20}{c}}
{{{| \alpha  |}^2}{M_0} + {{| \beta  |}^2}{M_{2}}}&{\alpha {\beta ^*}{M_0}}\\
{{\alpha ^*}\beta {M_0}}&{{{| \beta  |}^2}{M_0} + {{| \alpha  |}^2}{M_{1}}}
\end{array}} \right),  \label{rho2} \end{equation}
where
 ${M_0} \equiv \left( {1 - \gamma } \right)\sum\limits_n^{} {\frac{{{\gamma ^n}}}{{{Q_n}}}},$
${M_{1}} \equiv \left( {1 - \gamma } \right){| \mu  |^2}\sum\limits_n^{} {\frac{{n{\gamma ^n}}}{{{Q_n}}}},$
${M_{2}} \equiv  \left( {1 - \gamma } \right){| \mu  |^2}\sum\limits_n^{} {\frac{{\left( {n + 1} \right){\gamma ^n}}}{{{Q_n}}}}, $
satisfying $M_0+|\alpha|^2 M_1+|\beta|^2 M_2=1$.

The case without cosmic expansion corresponds to $\varepsilon =0$ and thus $\gamma =0$, then (\ref{rho2}) indeed reduces to (\ref{rho1}).

For $\varepsilon  \neq 0$,  even if  $\beta=0$, the final state $\rho$ of the qubit is a mixed state, in contrast with the case without cosmic expansion.

\begin{figure}[htb]
\centering
\includegraphics[width=0.8\textwidth]{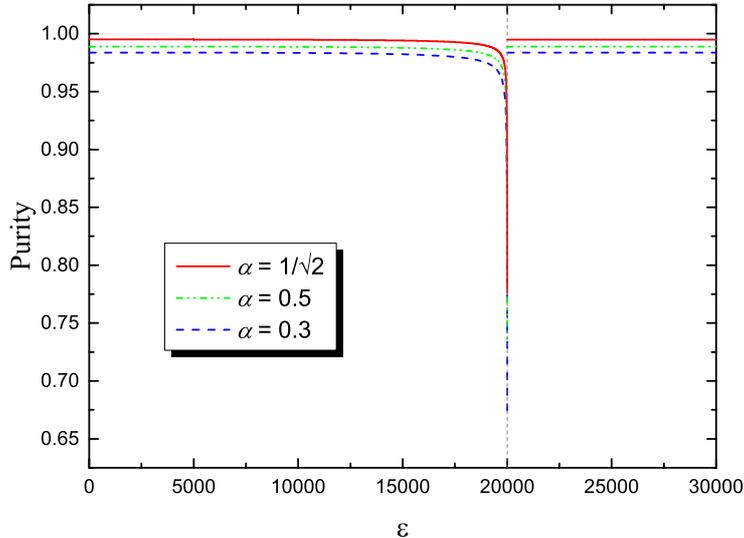}
\caption{The purity $\mathrm{Tr}\left( {\rho^2} \right)$ of the final state of the qubit as a function of the  volume $\varepsilon$ of the cosmic expansion. The expansion rapidity  is $\sigma=5$. The qubit energy level difference is $\Omega=2$, the inner product of the mode functions, defined in Eq.~(\ref{mu0}), is  $\mu  = 0.1$, the mass of the scalar field particle is  $m=0.01$.   The initial state of the qubit is $\alpha|0\rangle+\beta|1\rangle$.   } \label{fig1}
\end{figure}

\begin{figure}[htb]
\centering
\includegraphics[width=0.8\textwidth]{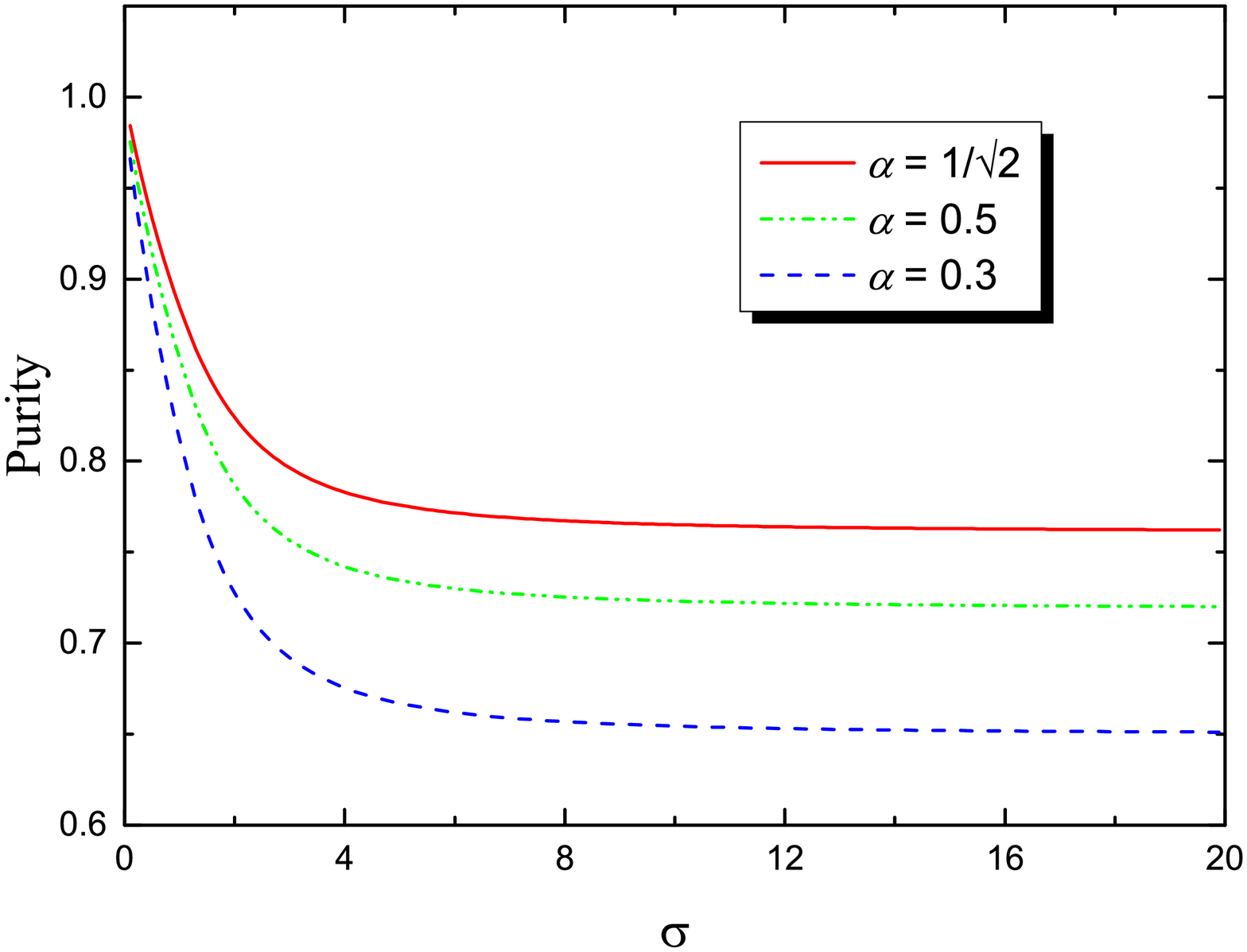}
\caption{The purity  $\mathrm{Tr}\left( {\rho^2} \right)$ of the final state of the qubit as a function of the expansion rapidity  $\sigma$ of the cosmic expansion. The qubit energy level difference is $\Omega=2$, the inner product of the mode functions, defined in Eq.~(\ref{mu0}), is $\mu  = 0.1$, the mass of the scalar field particle is $m=0.01$.  The volume  of the cosmic expansion is chosen to be $\varepsilon= {\varepsilon _{\max }} = 19999.5$.   The initial state of the qubit is $\alpha|0\rangle+\beta|1\rangle$. } \label{fig2}
\end{figure}

We calculate the purity $\mathrm{Tr}\left( {\rho^2} \right)$ of the qubit.   Its dependence   on the parameters $\varepsilon$ and $\sigma$ is as shown in Fig.~\ref{fig1}   and Fig.~\ref{fig2}.

\begin{figure}[htb]
\centering
\includegraphics[width=0.8\textwidth]{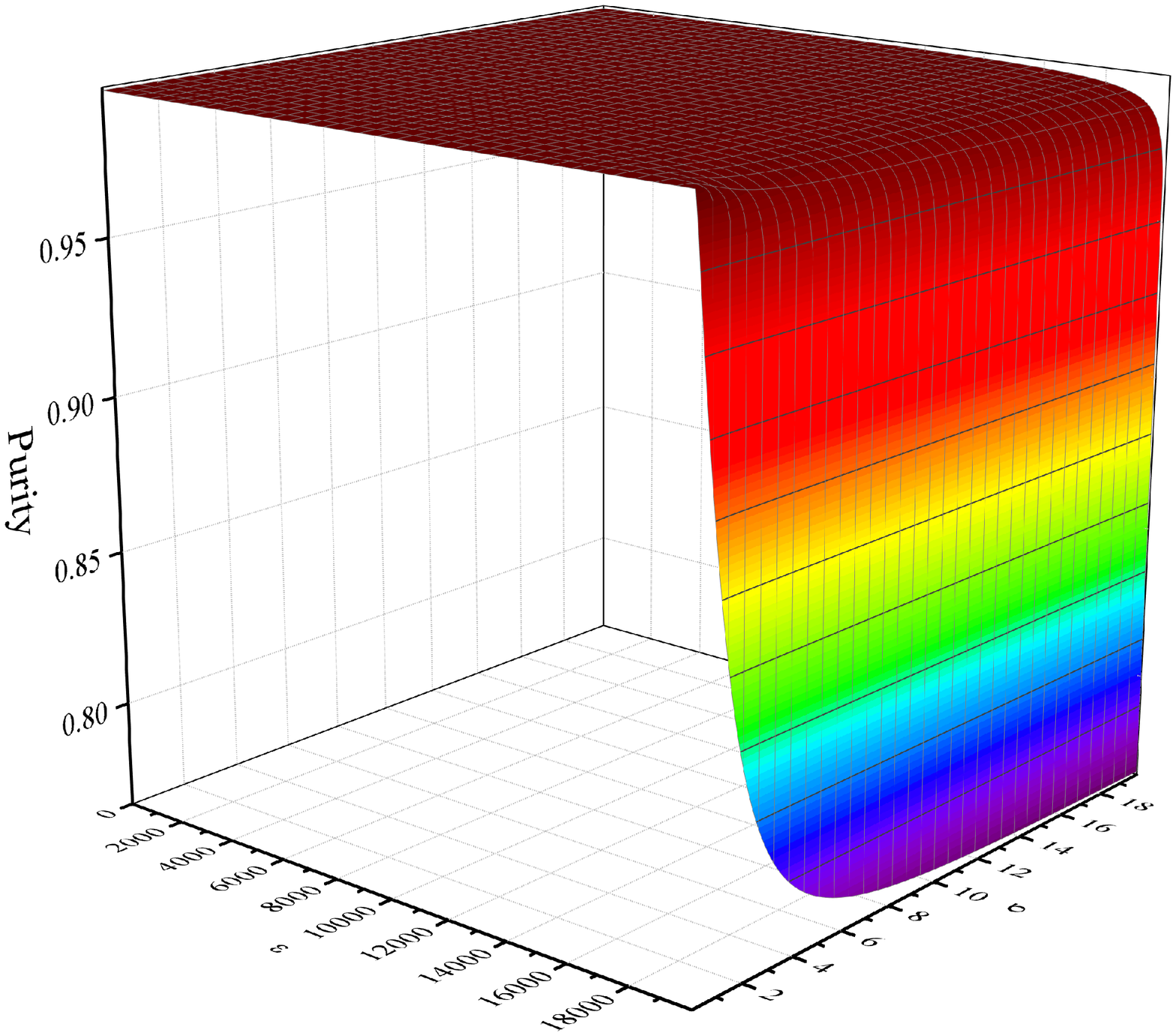}
\caption{The purity $\mathrm{Tr}\left( {\rho^2} \right)$ of the final state of the qubit as a function of the  volume $\varepsilon$ and  the rapidity $\sigma$ of the cosmic expansion.  The qubit energy level difference is $\Omega=2$, the inner product of the mode functions, defined in Eq.~(\ref{mu0}), is $\mu  = 0.1$, the mass of the scalar field particle is $m=0.01$.   The initial state of the qubit is $\frac{1}{\sqrt{2}}\left(|0\rangle+ |1\rangle\right)$.   } \label{fig3}
\end{figure}

As shown in Fig.~\ref{fig1}, the purity slowly decreases with the increase of  $\varepsilon$. When $\varepsilon$ is close to  $\varepsilon _{\max }$, the purity  decreases rapidly to the minimum.

For $\varepsilon  > {\varepsilon _{\max }}$, $\omega_{\rm{out}} > \Omega$ even if $k=0$, hence all   the field modes are decoupled with the qubit, and the purity of the qubit remains as the initial. In other words, in order to be decohered by the scalar field, the qubit must satisfy Eq.~(\ref{omega}). If there has not been expansion of the universe, the qubit must satisfy $\Omega \geq m$ in order to be coupled and thus decohered by the field.

The dependence of the purity on the expansion rapidity $\sigma$ is shown in Fig.~\ref{fig2}. When $\sigma$ is very small, the purity decreases rapidly with the increase of $\sigma$. Then it slowly approaches an asymptotic  value, which is dependent on $\varepsilon$ and the initial state.

With given values of $\varepsilon$ and $\sigma$, for  $|\alpha| \leq 1/\sqrt{2}$, the smaller $|\alpha|$, the smaller the purity.

For $\alpha=1/\sqrt{2}$, the 2D plot of the purity as a function of the two parameters is shown in Fig.~\ref{fig3}. The larger $\sigma$, the larger the rate of  decrease of the purity with respect to $\varepsilon $.   When $\sigma$ is small  enough, the dependence of the purity  on $\varepsilon$ is saturated.

\section{Two entangled qubits \label{two} }

Now we study the effect of the expansion of the universe on two  entangled qubits. Suppose  each qubit q = A, B interacts locally  with a scalar field $\Phi_q$, and does not interact with the scalar field around the other qubit, as the qubits are so far away from each other that there is no causal contact between one qubit on one hand, and the other qubit and its ambient field on the other.

For each qubit q and  its ambient field $\Phi_q$, the discussion in Sec.~\ref{single} applies. For qubit q, the energy difference between the two basis states $|1\rangle_q$ and $|0\rangle_q$ in the Schr\"{o}dinger picture is $\Omega_q$. As noted above, for qubit q, only the coupling with   the field mode $\mathbf{k}_{q0}$ with energy $\Omega_q$  needs to be considered.

Suppose the initial two-qubit entangled state is
     \begin{equation}
     | \Psi  \rangle  = \alpha |0 \rangle_A |1  \rangle_B  + \beta|1\rangle_A |0  \rangle_B, \label{psiab}
     \end{equation}
with ${| \alpha  |^2} + {| \beta  |^2} = 1$.  The entanglement in $|\Psi\rangle$ is quantified as the entanglement entropy $-|\alpha|^2 \log_2 |\alpha|^2 - \left(1-|\alpha|^2 \right) \log_2 \left(1-|\alpha|^2\right)$, hence is symmetric between $|\alpha|^2$ and $1-|\alpha|^2$, and is maximal when $|\alpha|=1/\sqrt{2}$. Without loss of generality, we will only consider examples with $|\alpha | \leq 1/\sqrt{2}$.

There having expansion of the universe, the initial state of the field modes  $\mathbf{k}_{A0} $,  $\mathbf{k}_{B0} $     and the two qubits is
 \begin{equation}
  \rho_{\mathbf{k}_{A0}} \otimes \rho_{\mathbf{k}_{B0}} \otimes | \Psi  \rangle \langle \Psi  |
  \end{equation}
 where
 \begin{equation}
 \rho _{\mathbf{k}_{q0}} = \left( 1 - \gamma_q\right)\sum_n \gamma _{q}^n |n\rangle_{q0}\langle n|  \end{equation}
is the density matrix of the field mode $\mathbf{k}_{q0}$, $n$ is the particle number in this mode.  After the interaction between each qubit and its ambient field, the density matrix of the qubits and the field modes evolves to
\begin{equation}
\left( 1 - \gamma_A \right)\left( 1 - \gamma_B\right) \sum_{n,m} \gamma_{A}^n\gamma_{B}^m |\phi_{nm}\rangle\langle \phi_{nm} |, \label{rhoentangle}
\end{equation}
where
\begin{equation}
\begin{array}{rl}
|\phi_{nm}\rangle \equiv & \frac{1}{\sqrt{Q_{nm}} } \{ \alpha \left(|n\rangle_{A0}|0\rangle_A  - i\sqrt n {\mu_A} |n - 1\rangle_{A0}|1\rangle_A   \right) \left(|m\rangle_{B0}|1\rangle_B  + i\sqrt {m + 1} \mu_B^*|m + 1\rangle_{B0}|0\rangle_B\right) \\& +
\beta \left(|n\rangle_{A0}|1 \rangle_A+ i\sqrt{n + 1} \mu_A^* | n + 1 \rangle_{A0}|0\rangle_A  \right) \left(|m\rangle_{B0}|0\rangle_B - i\sqrt {m } \mu_B|m - 1\rangle_{B0}|1\rangle_B\right) \},
\end{array}
\end{equation}
with
\begin{equation}
\begin{array}{rl}
Q_{nm} = &|\alpha|^2\left(1 + n|\mu_{A}|^2 + \left( m + 1 \right) |\mu_{B}|^2+ n\left( m + 1 \right)| \mu_{A}|^2|\mu_{B}|^2\right) \\& + |\beta|^2 \left(1 + m |\mu_{B}|^2 + \left( n + 1 \right) |\mu _{A}|^2  + m\left(n + 1\right) |\mu_{A}|^2 |\mu_{B}|^2\right). \end{array} \label{qnm}
\end{equation}
Tracing out the field modes yields  the final reduced density matrix of the qubits,
 \begin{equation}
 \begin{array}{rcl}
\rho^{AB} &=& \left( 1 - \gamma_A \right)\left(1 - \gamma_B \right) \displaystyle \sum_{n,m} \frac{\gamma_{A}^n\gamma_{B}^m}{
Q_{nm}} [ \left( | \alpha  |^2  + |\beta|^2 m\left( n + 1 \right) |\mu_{A}|^2|\mu_{B}|^2\right)
 |01\rangle \langle 01| \\
&& + \alpha\beta^*|01\rangle \langle 10| + \alpha^* \beta | 10 \rangle \langle 01|
   + \left(  | \alpha  |^2\left( {m + 1} \right) |  \mu _{B}  |^2  +|\beta|^2\left( {n + 1} \right)|\mu _{A}| ^2\right)|00\rangle \langle 00 | \\
    &&+ \left(  | \alpha  |^2 n |  \mu _{A}  | ^2  +  | \beta  | ^2 m | \mu _{B}  | ^2 \right)|  11  \rangle \langle {11} |   + \left(  | \alpha  | ^2 n\left( {m + 1} \right) | \mu _{A}  | ^2 | \mu _{B} | ^2  + | \beta  | ^2  \right) | 10  \rangle \langle  10 |  ],
\end{array}
\end{equation}
where $|ij\rangle \equiv | i \rangle_A| j \rangle_{B}$.
Using $| {00} \rangle $,  $| {01} \rangle $, $| {10} \rangle $, $| {11} \rangle $ as the basis states, the density matrix  can be written as
      \begin{equation}  {\rho^{AB}  } = \left( {\begin{array}{cccc}
 {{{| \alpha  |}^2}{S_4} + {{| \beta  |}^2}{S_3}} &0&0&0\\
0&{{{| \alpha  |}^2}{S_0} + {{| \beta  |}^2}{S_5}}&{\alpha {\beta ^*}{S_0}}&0\\
0&{{\alpha ^*}\beta {S_0}}&{{{| \alpha  |}^2}{S_6} + {{| \beta  |}^2}{S_{\rm{0}}}}&0\\
0&0&0&{{{| \alpha  |}^2}{S_1} + {{| \beta  |}^2}{S_2}}
\end{array}} \right),    \end{equation}
where
\begin{equation}  {S_0} \equiv \left( {1 - {\gamma _{A}}} \right)\left( {1 - {\gamma _{B}}} \right)\sum\limits_{n,m}^{} {\frac{{\gamma _{A}^n\gamma _{B}^m}}{{{Q_{nm}}}}},   \end{equation}
\begin{equation}{S_1}  \equiv \left( {1 - {\gamma _{A}}} \right)\left( {1 - {\gamma _{B}}} \right){| {{\mu _{A}}} |^2}\sum\limits_{n,m}^{} {\frac{{n\gamma _{A}^n\gamma _{B}^m}}{{{Q_{nm}}}}},    \end{equation}
\begin{equation}{S_2} \equiv\left( {1 - {\gamma _{A}}} \right)\left( {1 - {\gamma _{B}}} \right){| {{\mu _{B}}} |^2}\sum\limits_{n,m}^{} {\frac{{m\gamma _{A}^n\gamma _{B}^m}}{{{Q_{nm}}}}},       \end{equation}
\begin{equation} {S_3} \equiv\left( {1 - {\gamma _{A}}} \right)\left( {1 - {\gamma _{B}}} \right){| {{\mu _{A}}} |^2}\sum\limits_{n,m}^{} {\frac{{\left( {n + 1} \right)\gamma _{A}^n\gamma _{B}^m}}{{{Q_{nm}}}}},    \end{equation}
\begin{equation} S_4  \equiv \left( 1 - \gamma_{A} \right)\left(1 - \gamma_{B} \right) |\mu_{B} |^2 \sum\limits_{n,m} \frac{\left( m + 1 \right)\gamma_{A}^n\gamma_{B}^m}{Q_{nm}},      \end{equation}
\begin{equation} {S_5}  \equiv\left( {1 - {\gamma _{A}}} \right)\left( {1 - {\gamma _{B}}} \right){| {{\mu _{A}}} |^2}{| {{\mu _{B}}} |^2}\sum\limits_{n,m}^{} {\frac{{m\left( {n + 1} \right)\gamma _{A}^n\gamma _{B}^m}}{{{Q_{nm}}}}},    \end{equation}
\begin{equation} {S_6} = \left( {1 - {\gamma _{A}}} \right)\left( {1 - {\gamma _{B}}} \right){| {{\mu _{A}}} |^2}{| {{\mu _{B}}} |^2}\sum\limits_{n,m}^{} {\frac{{n\left( {m + 1} \right)\gamma _{A}^n\gamma _{B}^m}}{{{Q_{nm}}}}},     \end{equation}
satisfying $S_0+|\alpha|^2(S_1+S_4+S_6)+|\beta|^2(S_2+S_3+S_5)=1$.

The eigenvalues of $\rho^{AB}$ are
       \begin{equation}  {\lambda^{AB} _1} = {| \alpha  |^2}{S_4} + {| \beta  |^2}{S_3},  \end{equation}
        \begin{equation} {\lambda^{AB} _2} = {| \alpha  |^2}{S_1} + {| \beta  |^2}{S_2},    \end{equation}

\begin{equation}
\begin{array}{rl}
\lambda^{AB}_{3,4} = & \frac{1}{2} \left[ S_0 + |\alpha|^2 S_6 + |\beta|^2 S_5 \pm \right. \\
 & \left. \sqrt{S_0^2 + ( | \alpha  |^2 S_6 - |\beta|^2 S_5 )^2 + 2(|\alpha|^2 -
|\beta|^2 ) S_0 (|\beta|^2 S_5 - |\alpha|^2 S_6 ) } \right].
\end{array}
\end{equation}

The reduced density matrix of A is
      \begin{equation} \rho^{A} = \left( {\begin{array}{*{20}{c}}
{{{| \alpha  |}^2}\left( {{S_0} + {S_4}} \right) + {{| \beta  |}^2}\left( {{S_3} + {S_5}} \right)}&0\\
0&{{{| \alpha  |}^2}\left( {{S_1} + {S_6}} \right) + {{| \beta  |}^2}\left( {{S_0} + {S_2}} \right)}
\end{array}} \right),  \end{equation}
with eigenvalues
       \begin{equation} \lambda _1^{A} = {| \alpha  |^2}\left( {{S_0} + {S_4}} \right) + {| \beta  |^2}\left( {{S_3} + {S_5}} \right),   \end{equation}
        \begin{equation}  \lambda _2^{A} = {| \alpha  |^2}\left( {{S_1} + {S_6}} \right) + {| \beta  |^2}\left( {{S_0} + {S_2}} \right).   \end{equation}

The reduced density matrix of B   is
\begin{equation}\rho^{B} = \left( {\begin{array}{*{20}{c}}
{{{| \alpha  |}^2}\left( {{S_4} + {S_6}} \right) + {{| \beta  |}^2}\left( {{S_0} + {S_3}} \right)}&0\\
0&{{{| \alpha  |}^2}\left( {{S_0} + {S_1}} \right) + {{| \beta  |}^2}\left( {{S_2} + {S_5}} \right)}
\end{array}} \right), \end{equation}
with eigenvalues
     \begin{equation}  \lambda _1^{B} = {| \alpha  |^2}\left( {{S_4} + {S_6}} \right) + {| \beta  |^2}\left( {{S_0} + {S_3}} \right),   \end{equation}
      \begin{equation} \lambda _2^{B} = {| \alpha  |^2}\left( {{S_0} + {S_1}} \right) + {| \beta  |^2}\left( {{S_2} + {S_5}} \right).      \end{equation}

We now study the mutual information
\begin{equation}
I = S\left( {\rho^{A}} \right) + S\left( {\rho^{B}} \right) - S\left( {{\rho^{AB}  }} \right),\end{equation}
 which is a quantifying measure of the total correlation, contributed by both quantum entanglement and classical correlation. It can be obtained as
       \begin{equation} \begin{array}{rcl}
I & = & - \lambda _1^{A}{\log _2}\lambda _1^{A} - \lambda _2^{A}{\log _2}\lambda _2^{A} - \lambda _1^{B}{\log _2}\lambda _1^{B} - \lambda _2^{B}{\log _2}\lambda _2^{B}\\
&& + {\lambda^{AB} _1}{\log _2}{\lambda^{AB} _1} + {\lambda^{AB} _2}{\log _2}{\lambda^{AB} _2} + {\lambda^{AB} _3}{\log _2}{\lambda^{AB} _3} + {\lambda^{AB} _4}{\log _2}{\lambda^{AB} _4}.
\end{array}  \end{equation}

\begin{figure}[htb]
\centering
\includegraphics[width=0.8\textwidth]{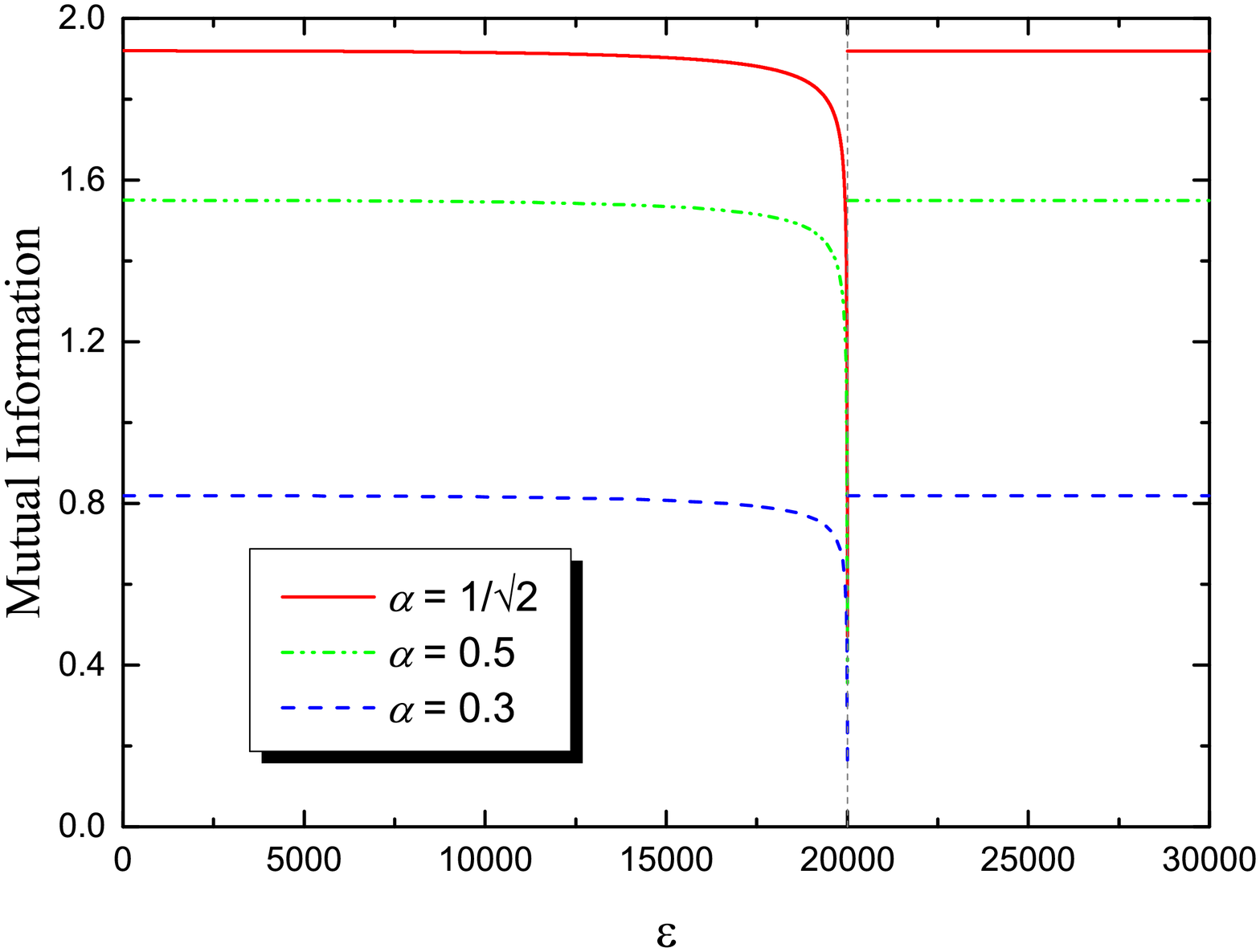}
\caption{ The mutual information $I$ of the final state of  two entangled qubits as a function of the  volume $\varepsilon$ of the cosmic expansion. The expansion rapidity  is $\sigma=5$. The parameters of the two qubits are the same. The qubit energy level difference is $\Omega=2$, the inner product of the mode functions, defined in Eq.~(\ref{mu0}), is $\mu  = 0.1$, the mass of the scalar field particle is $m=0.01$.  The initial state of the qubits is $\alpha|01\rangle+\beta|10\rangle$.  } \label{fig4}
\end{figure}

\begin{figure}[htb]
\centering
\includegraphics[width=0.8\textwidth]{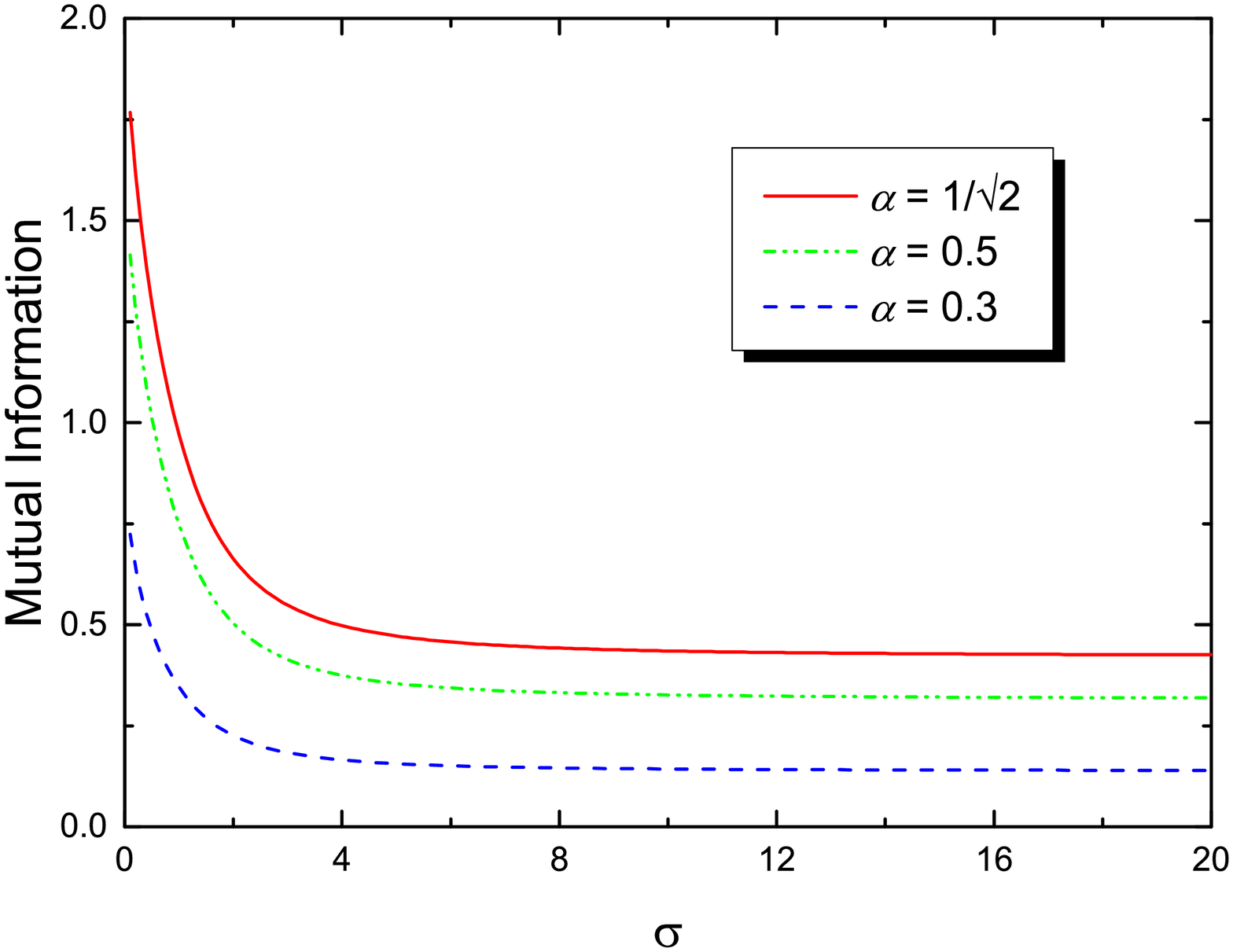}
\caption{The mutual information $I$ of the final state of  two entangled qubits as a function of the rapidity   $\sigma$ of the expansion of the universe. The parameters of the two qubits are the same. The qubit energy level difference is $\Omega=2$, the inner product of the mode functions, defined in Eq.~(\ref{mu0}), is $\mu  = 0.1$, the mass of the scalar field particle is $m=0.01$. The volume of the expansion is chosen to be   ${\varepsilon  =\varepsilon _{\max }} = 19999.5$.  The initial state of the qubits is $\alpha|01\rangle+\beta|10\rangle$.  } \label{fig5}
\end{figure}

\begin{figure}[htb]
\centering
\includegraphics[width=0.8\textwidth]{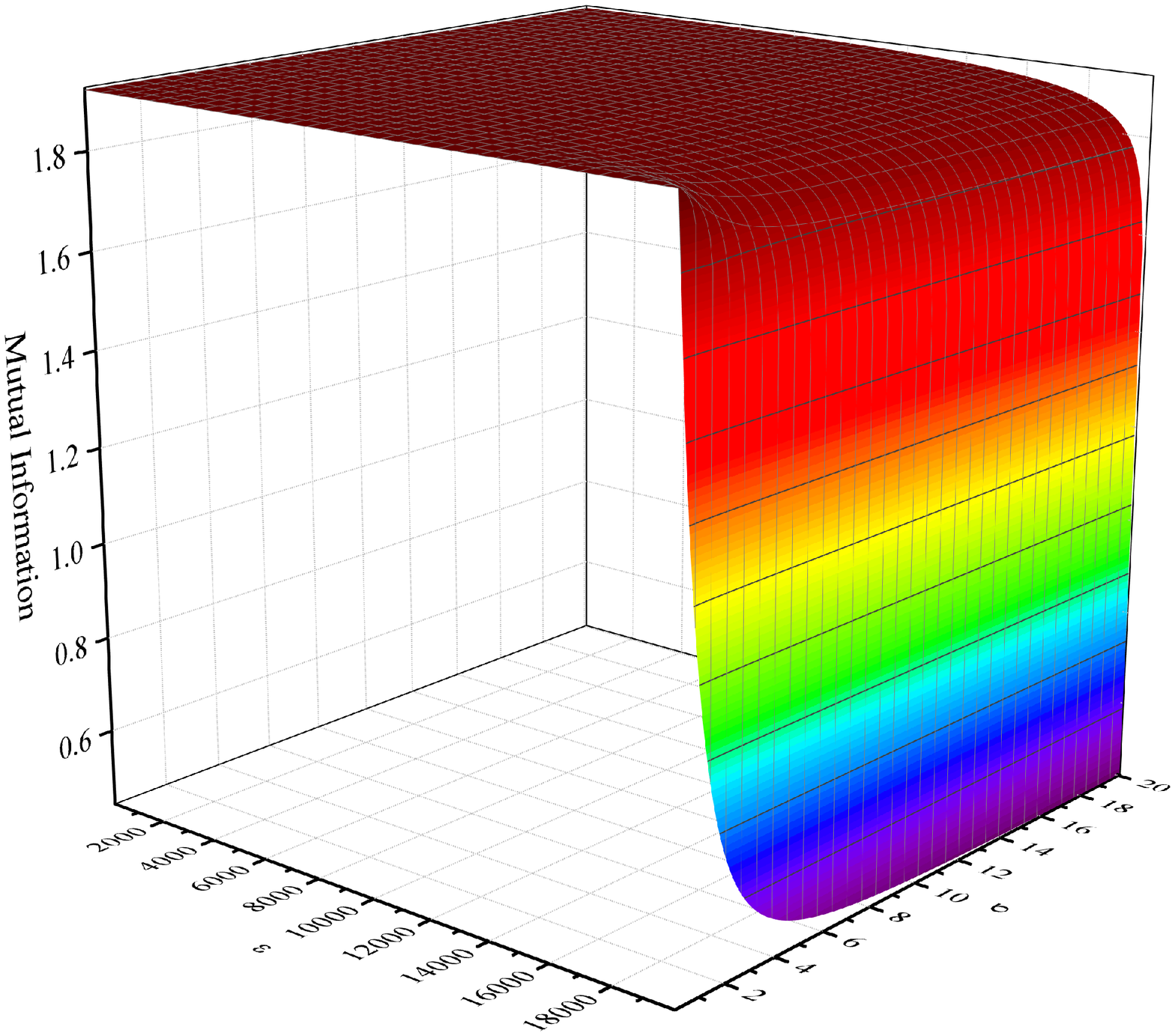}
\caption{The mutual information $I$ of the final state of  two entangled qubits as a function of the  volume $\varepsilon$ and  the rapidity $\sigma$ of the cosmic expansion.   The parameters of the two qubits are the same. The qubit energy level difference is $\Omega=2$, the inner product of the mode functions, defined in Eq.~(\ref{mu0}), is $\mu  = 0.1$, the mass of the scalar field particle is $m=0.01$.   The initial state of the qubits is  $\frac{1}{\sqrt{2}}\left(|01\rangle+ |10\rangle\right)$.} \label{fig6}
\end{figure}

The dependence of the mutual information  $I$ on the parameters $\varepsilon$ and $\sigma$ is shown in Figs.~\ref{fig4} and \ref{fig5}, respectively. $\varepsilon =0$  corresponds to the case that there has not been expansion of the universe.  It can be seen that for a given value of  $\sigma$, as far as $0 < \varepsilon \leq {\varepsilon _{\max }}$, the mutual information $I$ monotonically decreases with  the increase of $\varepsilon$, and decreases more rapidly when $\varepsilon$ is closer to  ${\varepsilon _{\max }}$.   For a given value of  $0< \varepsilon \leq {\varepsilon _{\max }}$,  the mutual information also monotonically decreases with the increase of  $\sigma$.  The larger $\sigma$, the smaller the rate of the decrease, and the mutual information approaches an asymptotic value as $\sigma$ increases.  Moreover, for $|\alpha| \leq 1/\sqrt{2}$, the smaller $|\alpha|$, i.e. the smaller the initial entanglement, the smaller the mutual information at  given values of  $\varepsilon$ and $\sigma$.

For $\alpha=1/\sqrt{2}$, the 2D plot of the mutual information   as a function of the two parameters $\varepsilon$ and $\sigma$ is shown in Fig.~\ref{fig6}. The larger $\sigma$, the larger the rate of decrease of the mutual information with respect to $\varepsilon $. When $\sigma$ is small enough, the dependence   on $\varepsilon$ is saturated.

Now we study how the entanglement between the two qubits is degraded in an expanded  universe. The entanglement between qubits A and B in a mixed state is quantified  by the concurrence~\cite{wootters}
 \begin{equation}
 C\left( \rho^{AB}  \right) = \max \left\{ {0,\sqrt {{\lambda _1}}  - \sqrt {{\lambda _2}}  - \sqrt {{\lambda _3}}  - \sqrt {{\lambda _4}} } \right\},    \end{equation}
where ${\lambda _i}$'s are  the eigenvalues of
\begin{equation}
X\equiv {\rho ^{{AB}}}\left( {{\sigma _y} \otimes {\sigma _y}} \right) \left(\rho ^{AB}\right)^* \left( {{\sigma _y} \otimes {\sigma _y}} \right),
\end{equation}
satisfying ${\lambda _1} \ge {\lambda _2} \ge {\lambda _3} \ge {\lambda _4}$, $\left(\rho ^{AB}\right)^*$ is the complex conjugate of ${\rho ^{{AB}}}$, ${\sigma _y} = \left( {\begin{array}{*{20}{c}}
0&{ - i}\\
i&0
\end{array}} \right)$.

We obtain
\begin{equation}
X = \left( {\begin{array}{cccc}
W&0&0&0 \\
0&V&Y&0\\
0&Z&V&0\\
0&0&0&W
\end{array}} \right),  \end{equation}
where
\begin{equation} V \equiv {| \alpha  |^4}{S_0}{S_6} + {| \beta  |^4}{S_0}{S_5} + {| \alpha  |^2}{| \beta  |^2}\left( {2S_0^2 + {S_5}{S_6}} \right),    \end{equation}
\begin{equation} W \equiv  {| \alpha  |^4}{S_1}{S_4} + {| \beta  |^4}{S_2}{S_3} + {| \alpha  |^2}{| \beta  |^2}\left( {{S_2}{S_4} + {S_1}{S_3}} \right),     \end{equation}
\begin{equation}
 Y \equiv 2\alpha {\beta ^*}\left( {{{| \alpha  |}^2}{S_0} + {{| \beta  |}^2}{S_5}} \right){S_0},   \end{equation}
\begin{equation}
 Z\equiv  2{\alpha ^*}\beta \left( {{{| \alpha  |}^2}{S_6} + {{| \beta  |}^2}{S_0}} \right){S_0}.    \end{equation}

The eigenvalues of $X$  can be obtained as
\begin{equation} \lambda _{1,2} = {| \alpha  |^4}{S_1}{S_4} + {| \beta  |^4}{S_2}{S_3} + {| \alpha  |^2}{| \beta  |^2}\left( {{S_2}{S_4} + {S_1}{S_3}} \right),    \end{equation}
\begin{equation} \lambda _{3,4} = \left[ {| \alpha  || \beta  |{S_0} \pm \sqrt {\left( {{{| \alpha  |}^2}{S_0} + {{| \beta  |}^2}{S_5}} \right)\left( {{{| \alpha  |}^2}{S_6} + {{| \beta  |}^2}{S_0}} \right)} } \right]^2,    \end{equation}
using which the concurrence is calculated,  as a function of $\varepsilon$ and $\sigma$, respectively.

\begin{figure}[htb]
\centering
\includegraphics[width=0.8\textwidth]{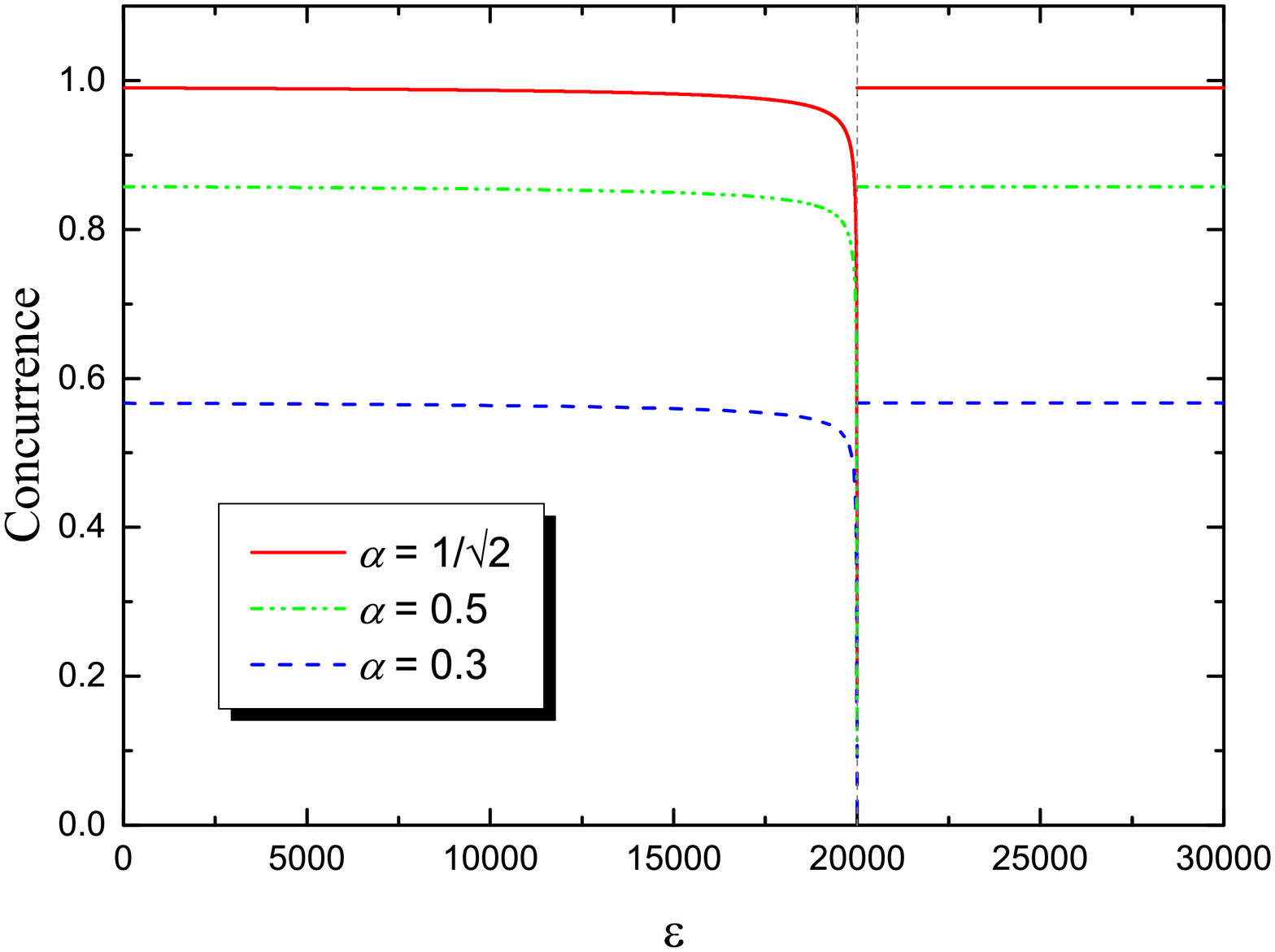}
\caption{The concurrence  of the final state of  two entangled qubits as a function of the  volume $\varepsilon$ of the cosmic expansion. The expansion rapidity is $\sigma=5$. The parameters of the two qubits are the same. The qubit energy level difference is $\Omega=2$, the inner product of the mode functions, defined in Eq.~(\ref{mu0}), is $\mu  = 0.1$, the mass of the scalar field particle is $m=0.01$.  The initial state of the qubits is $\alpha|01\rangle+\beta|10\rangle$.
} \label{fig7}
\end{figure}

\begin{figure}[htb]
\centering
\includegraphics[width=0.8\textwidth]{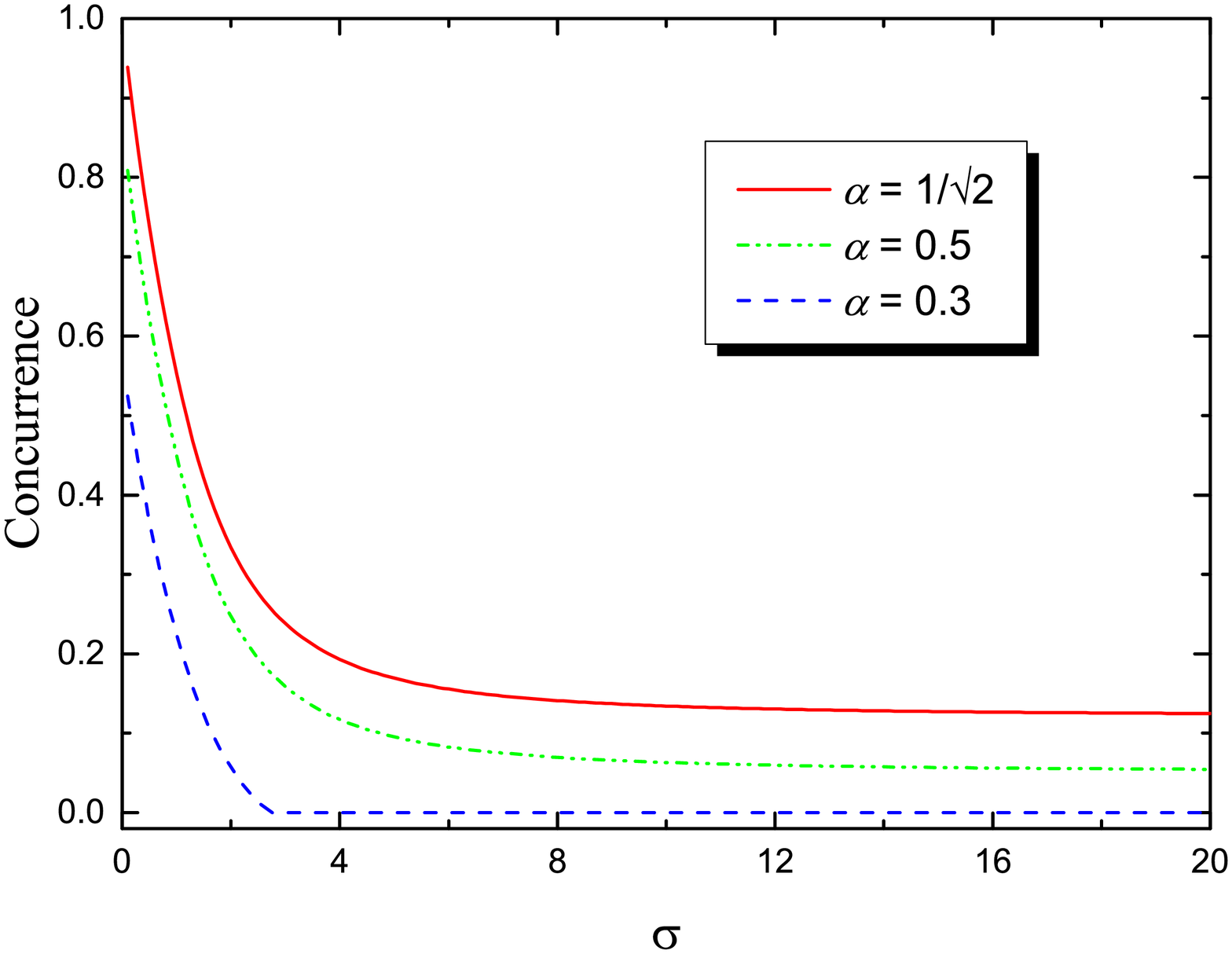}
\caption{The concurrence  of the final state of two entangled qubits as a function of the rapidity   $\sigma$ of the expansion of the universe. The parameters of the two qubits are the same. The qubit energy level difference is $\Omega=2$, the inner product of the mode functions, defined in Eq.~(\ref{mu0}), is $\mu  = 0.1$, the mass of the scalar field particle is $m=0.01$. The volume of the expansion is chosen to be ${\varepsilon  =\varepsilon _{\max }} = 19999.5$.  The initial state of the qubits is $\alpha|01\rangle+\beta|10\rangle$.} \label{fig8}
\end{figure}

\begin{figure}[htb]
\centering
\includegraphics[width=0.8\textwidth]{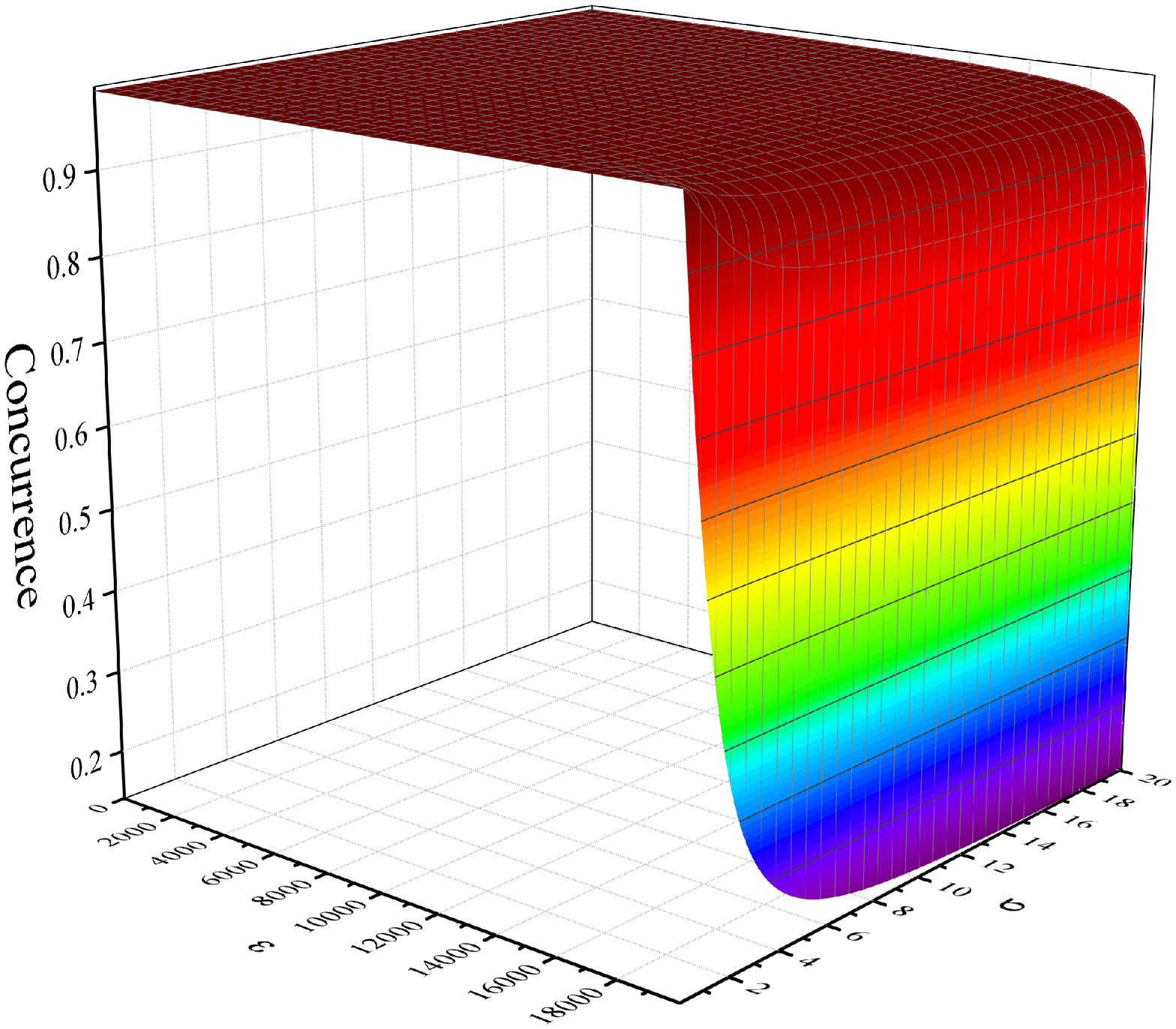}
\caption{The concurrence  of the final state of  two entangled qubits as a function of the  volume $\varepsilon$ and  the rapidity $\sigma$ of the cosmic expansion.  The parameters of the two qubits are the same. The qubit energy level difference is $\Omega=2$, the inner product of the mode functions, defined in Eq.~(\ref{mu0}), is $\mu  = 0.1$, the mass of the scalar field particle is $m=0.01$.   The initial state of the two qubits is  $\frac{1}{\sqrt{2}}\left(|01\rangle+ |10\rangle\right)$. } \label{fig9}
\end{figure}

As  shown in Fig.~\ref{fig7},   the concurrence monotonically decreases with the increase of $\varepsilon$,  reaching the   minimum at ${\varepsilon _{\max }} $.  As  shown in Fig.~\ref{fig8},
the concurrence also decreases with the increase of $\sigma$. The larger $\sigma$, the smaller the rate of the decrease, and   the concurrence approaches an asymptotic value as $\sigma$ increases.

Moreover,  with given values of  $\varepsilon$ and $\sigma$, for $|\alpha| \leq 1/\sqrt{2}$ and larger than a certain value, the smaller $\alpha$, i.e. the smaller the initial entanglement, the smaller the final concurrence. A significant feature is that when $|\alpha|$ is smaller than a certain value, i.e. when the initial entanglement is not too large, there exists entanglement sudden death~\cite{yu}  at a finite value of $\sigma$, that is, for  $\sigma$ larger than this critical  value, the entanglement remains vanishing.

For $\alpha=1/\sqrt{2}$, the 2D plot of the concurrence as a function of the two parameters is shown in Fig.~\ref{fig9}. The larger $\sigma$, the larger the rate of decrease of  the concurrence with respect to $\varepsilon $. When $\sigma$ is small enough, the dependence of  the concurrence on $\varepsilon$ is saturated.

If there has not been expansion of the universe, the initial states of the field modes are vacua, hence the evolution of the  state of the field modes and the qubits is
\begin{equation}
| 0 \rangle_{\mathbf{k}_{A0} }   \otimes | 0 \rangle_{\mathbf{k}_{B0} }   \otimes   |\Psi\rangle
 \rightarrow
 |0\rangle_{\mathbf{k}_{A0} } |0\rangle_{\mathbf{k}_{B0} }|\Psi\rangle
  +i\left(\alpha\mu_B^*  |0\rangle_{\mathbf{k}_{A0} } |1\rangle_{\mathbf{k}_{B0} }
  +\beta \mu_A^* |1\rangle_{\mathbf{k}_{A0} } |0\rangle_{\mathbf{k}_{B0} }\right)  |00\rangle.
\end{equation}
Hence the reduced density matrix of the qubits is
\begin{eqnarray}
\rho^{AB} & = & \frac{1}{1+|\alpha|^2|\mu_B|^2+|\beta|^2|\mu_A|^2} \left[ |\Psi\rangle\langle \Psi| + (|\alpha|^2|\mu_B|^2+|\beta|^2|\mu_A|^2)|00\rangle\langle00|\right] \\& =  & \frac{1}{1+|\alpha|^2|\mu_B|^2+|\beta|^2|\mu_A|^2} \left( {\begin{array}{cccc}
 |\alpha|^2|\mu_B|^2+|\beta|^2|\mu_A|^2  &0&0&0\\
0& | \alpha  |^2 & \alpha {\beta ^*} &0\\
0& \alpha ^* \beta  & | \beta  |^2 &0\\
0&0&0&0
\end{array}} \right),    \end{eqnarray}
which is a mixed state even if $\alpha=0$ or $\beta=0$.

It can be obtained that in this case, the mutual information is
\begin{equation}
\begin{array}{rl}
I = &\log_2(1+|\alpha|^2|\mu_B|^2+|\beta|^2|\mu_A|^2)
-\frac{1}{1+|\alpha|^2|\mu_B|^2+|\beta|^2|\mu_A|^2}[|\alpha|^2\log_2 |\alpha|^2
+|\beta|^2\log_2 |\beta|^2\\
&+(|\alpha|^2+|\alpha|^2|\mu_B|^2+|\beta|^2|\mu_A|^2) \log_2 (|\alpha|^2+|\alpha|^2|\mu_B|^2+|\beta|^2|\mu_A|^2) \\ & + (|\beta|^2+|\alpha|^2|\mu_B|^2+|\beta|^2|\mu_A|^2) \log_2 (|\beta|^2+|\alpha|^2|\mu_B|^2+|\beta|^2|\mu_A|^2)\\&- (|\alpha|^2|\mu_B|^2+|\beta|^2|\mu_A|^2) \log_2 (|\alpha|^2|\mu_B|^2+|\beta|^2|\mu_A|^2) ],
\end{array}
\end{equation}
which vanishes when $\alpha=0$ or $\beta =0$, and is nonzero if $\alpha \neq 0$ and $\beta \neq 0$. Moreover, without the coupling with the fields, $\mu_A = \mu_B=0$, the mutual information reduces to $I=-2|\alpha|^2 \log_2 |\alpha|^2 - 2 |\beta|^2 \log_2 |\beta|^2$, which is that of the original state $| \Psi  \rangle  = \alpha |0 \rangle_A |1  \rangle_B  + \beta|1\rangle_A |0  \rangle_B$.

In this case,  the concurrence is
\begin{equation}
C(\rho^{AB}) = \frac{2|\alpha||\beta|}{1+|\alpha|^2|\mu_B|^2+|\beta|^2|\mu_A|^2},
\end{equation}
implying that in the case that there has not expansion of the universe, if and only if the initial state is an entangled state, i.e. $\alpha\neq 0$ and $\beta \neq 0$, the final state is entangled. Moreover,   without the coupling with the fields, $\mu_A = \mu_B=0$, the concurrence reduces to $2|\alpha||\beta|$, which is that of the original state $| \Psi  \rangle  = \alpha |0 \rangle_A |1  \rangle_B  + \beta|1\rangle_A |0  \rangle_B$.

\section{Quantum teleportation \label{teleportation} }

Another way of characterizing the entanglement degradation of a maximally entangled state  is in terms of the fidelity of  quantum teleportation.  Quantum teleportation is a quantum information protocol based  on a maximally entangled state~\cite{bennett}. In the ideal case, without environmental disturbance,  two qubits initially entangled remain entangled no matter how far they are separated. The coupling with a scalar field causes the qubits entangled with the field modes, hence the entanglement between the two qubits degrades. Consequently the teleportation based on the two-qubit entanglement is disturbed. The fidelity of the teleportation measures how well the teleportation is completed in presence of the coupling with a field. As we have seen in the last section, the entanglement degradation is greatly enhanced by the expansion of the universe. Hence the teleportation fidelity provides a witness of the cosmic expansion.

Consider in the far future a two-qubit state, which was initially a maximally entangled state but has been degraded by the scalar field. We study how the teleportation fidelity depends on the parameters of the cosmic expansion.

Suppose the two qubits A and B are prepared to be  one of the four Bell states
\begin{equation}  \begin{array}{l}
| {{\Phi ^\pm }} \rangle  = \frac{1}{{\sqrt 2 }}\left( {| 0 \rangle | 0 \rangle  \pm | 1 \rangle | 1 \rangle } \right),\\
| {{\Psi ^ \pm }} \rangle  = \frac{1}{{\sqrt 2 }}\left( {| 0 \rangle | 1 \rangle  \pm | 1 \rangle | 0 \rangle } \right),
\end{array} \end{equation}
say, $|\Psi^+\rangle$.

A and B are  separated at two locations far away from each other. Each of the two qubits interacts locally with the scalar field around it. Hence the state of the two qubits together with the two relevant field modes is given by Eq.~(\ref{rhoentangle}), with $\alpha=\beta=1/\sqrt{2}$.

As in the usual protocol of teleportation, A and B are respectively controlled by Alice and Bob. Alice also controls another qubit C, which is   in the state
\begin{equation}
|\psi\rangle_C = u |0\rangle_C+ v |1\rangle_C.
\end{equation}

Our interest lies in the teleportation fidelity decrease due to the effect of scalar field on   the resource, namely the entangled state $|\Psi^+\rangle$, which is shared by two qubits far away from each other.  The teleportation can start immediately after $|\psi\rangle_C $ is prepared, which is thus not considered to be affected by the scalar field.

The composite  state of the three qubits and the two field modes is
\begin{equation}
\rho^{\rm{all}}=
\left( 1 - \gamma_A \right)\left( 1 - \gamma_B\right) \sum_{n,m} \gamma_{A}^n\gamma_{B}^m |\Phi_{nm},\psi\rangle \langle\Phi_{nm},\psi|, \label{rhoentanglec}
\end{equation}
where
\begin{equation}
\begin{array}{rl}
&|\Phi_{nm},\psi\rangle  \\
&\equiv  \frac{1}{ 2\sqrt{Q_{nm}}}
 \{ [\left(u | n \rangle_{A0}- iv \sqrt {n} \mu_{A}|n-1 \rangle_{A0}\right)\left( | m\rangle  _{B0} |1\rangle_B+ i\sqrt {m + 1} \mu_{B}^*| m+1\rangle _{B0} |0\rangle_B\right)\\
 & +\left(v | n \rangle_{A0}+ iu \sqrt {n+1} \mu_{A}^*|n + 1 \rangle_{A0}\right)\left( | m\rangle  _{B0} |0\rangle_B- i\sqrt {m} \mu_{B}| m-1\rangle_{B0} |1\rangle_B\right)]| \Phi^+ \rangle_{AC}\\
  &+[\left(u | n \rangle_{A0} + iv \sqrt {n} \mu_{A}|n-1 \rangle_{A0}\right)\left( | m\rangle _{B0} |1\rangle_B+ i\sqrt {m + 1} \mu_{B}^*| m+1\rangle_{B0} |0\rangle_B\right)\\
  &+\left(- v | n \rangle_{A0}+ iu \sqrt {n+1} \mu_{A}^*|n + 1 \rangle_{A0}\right)\left( | m\rangle _{B0} |0\rangle_B- i\sqrt {m} \mu_{B}| m-1\rangle_{B0} |1\rangle_B\right)]| \Phi^- \rangle_{AC}\\
 &+ [\left(v | n \rangle_{A0}- iu \sqrt {n} \mu_{A}|n-1 \rangle_{A0}\right)\left( | m \rangle _{B0} |1\rangle_B+ i\sqrt {m + 1} \mu_{B}^*| m + 1\rangle _{B0} |0\rangle_B \right)\\
  &+\left(u | n \rangle_{A0}+ iv \sqrt {n+1} \mu_{A}^*|n + 1 \rangle_{A0}\right)\left( | m\rangle _{B0} |0\rangle_B- i\sqrt {m} \mu_{B}| m-1\rangle_{B0} |1\rangle_B\right)]| \Psi^+ \rangle_{AC}\\
 &+[\left(v | n \rangle_{A0}+ i u \sqrt {n} \mu_{A}|n-1 \rangle_{A0}\right)\left( | m\rangle_{B0} |1\rangle_B+ i\sqrt {m + 1} \mu_{B}^*| m + 1\rangle _{B0}|0\rangle_B \right)\\
  &+\left(- u | n \rangle_{A0}+ iv \sqrt {n+1} \mu_{A}^*|n + 1 \rangle_{A0}\right)\left( | m\rangle _{B0} |0\rangle_B- i\sqrt {m} \mu_{B}| m-1\rangle_{B0} |1\rangle_B\right)]| \Psi^- \rangle_{AC} \},
\end{array}
\label{psiphic}
\end{equation}
where $Q_{nm}$ is given by (\ref{qnm}) with $\alpha = \beta =1/\sqrt{2}$. For convenience, one can write
\begin{equation}
|\Phi_{nm},\psi\rangle   \equiv  \frac{1}{ 2\sqrt{Q_{nm}}} \sum_{i=1}^4 |\Phi^i_{nm}\rangle   |{\rm{Bell}}^i\rangle_{AC},
\end{equation}
where the terms  $i=1,2,3,4$ referrs to the four terms consecutively, $|{\rm{Bell}}^1\rangle=| \Phi^+ \rangle $,  $|{\rm{Bell}}^2\rangle=| \Phi^- \rangle $,  $|{\rm{Bell}}^3\rangle=| \Psi^+ \rangle $, $|{\rm{Bell}}^4\rangle=| \Psi^-\rangle$,  $|\Phi^i_{nm}\rangle $ is not normalized, and one can find
\begin{equation}\begin{array}{rl}
& \langle \Phi^1_{nm}|\Phi^1_{nm}\rangle    = \langle \Phi^2_{nm} |\Phi^2_{nm}\rangle = Q_{nm}^{v u } \equiv 1 + [ {{{| v |}^2}n + {{| u  |}^2}\left( {n + 1} \right)} ]{| {{\mu _{A}}} |^2} \\
 & +  [ {{{|v  |}^2}m + {{| u  |}^2}\left( {m + 1} \right)} ]{| {{\mu _{B}}} |^2}+ [ {{{|v  |}^2}n\left( {m + 1} \right) + {{| u  |}^2}\left( {n + 1} \right)m} ]{| {{\mu _{A}}} |^2}{| {{\mu _{B}}} |^2},
\end{array}       \end{equation}
\begin{equation}\begin{array}{rl}
 & \langle \Phi^3_{nm}|\Phi^3_{nm}\rangle=  \langle \Phi^4_{nm}|\Phi^4_{nm}\rangle = Q_{nm}^{u v }\equiv 1 + [ {{{| u  |}^2}n + {{| v  |}^2}\left( {n + 1} \right)} ]{| {{\mu _{A}}} |^2} \\
 &+ [ {{{| u  |}^2}m + {{| v  |}^2}\left( {m + 1} \right)} ]{| {{\mu _{B}}} |^2}+ [ {{{| u  |}^2}n\left( {m + 1} \right) + {{| v  |}^2}\left( {n + 1} \right)m} ]{| {{\mu _{A}}} |^2}{| {{\mu _{B}}} |^2},
\end{array}       \end{equation}
satisfying $Q_{nm}^{v u }+Q_{nm}^{u v }=2Q_{nm}$. By setting $\mu_A=\mu_B=0$, $\gamma =0$, and keeping only $n=m=0$ term, (\ref{rhoentanglec}) reduces to the ideal case of a pure state.

On qubits A and C, Alice makes a Bell measurement, i.e. a measurement in the basis of Bell states. In the ideal quantum teleportation,  the state of qubit B is a pure state.  After Alice informs Bob her measurement result  through classical communication, Bob  can  transform the state of qubit B to $|\psi\rangle$, which was  the original  state of C,  by using a one-qubit unitary transformation,  with a one-to-one correspondence with the four Bell states of qubits A and C. If the result of Bell measurement is $|\Phi^+\rangle$, then the state of qubit B  is $u |1\rangle+ v |0\rangle$, which can be transformed to $|\psi\rangle$ by $\sigma_x$. If the result of Bell measurement is $|\Phi^-\rangle$, then the state of qubit B is $u |1\rangle- v |0\rangle$, which can be transformed to $|\psi\rangle$ by $i\sigma_y$. If the result of Bell measurement is $|\Psi^+\rangle$, then the state of qubit  B is $u |0\rangle+ v |1\rangle$, which is just $|\psi\rangle$. If the result of Bell measurement is $|\Psi^-\rangle$, then the state of qubit B is $u |0\rangle- v |1\rangle$, which can be transformed to $|\psi\rangle$ by $\sigma_z$.

In presence  of the coupling with the field modes, the procedure of the teleportation remains the same. Although  usually $|\Phi^i_{nm}\rangle $'s are not orthogonal to each other, Bell states are orthogonal to each other, therefore  after Bell measurement of qubits A and C, each $|\Phi_{nm},\psi\rangle$ does collapse  into one of the four terms in (\ref{psiphic}), i.e.  $|\Phi^i_{nm}\rangle   |{\rm{Bell}}^i\rangle$, with the Bell state of qubits A and C  disentangled with the state of  qubit B and the field modes. Hence if qubits A and C are measured to be in  $|{\rm{Bell}}^i\rangle$, the density matrix of qubit B and the field modes becomes
\begin{equation}
\rho_{i}=
\left( 1 - \gamma_A \right)\left( 1 - \gamma_B\right) \sum_{n,m} \gamma_{A}^n\gamma_{B}^m \frac{|\Phi^i_{nm}\rangle \langle\Phi^i_{nm}|}{\langle \Phi^i_{nm}|\Phi^i_{nm}\rangle}. \label{rhobmodes}
\end{equation}
The probability for the result of the Bell measurement to be $|{\rm{Bell}}^i\rangle$ and thus $\rho^{all}$ collapses to $\rho_i\otimes|{\rm{Bell}}^i\rangle\langle{\rm{Bell}}^i|$ is
\begin{equation}
p_i=
\left( 1 - \gamma_A \right)\left( 1 - \gamma_B\right) \sum_{n,m} \gamma_{A}^n\gamma_{B}^m \frac{\langle \Phi^i_{nm}|\Phi^i_{nm}\rangle}{4Q_{nm}}.
\end{equation}
Thus
\begin{eqnarray}
p_1=p_2 =&
\left( 1 - \gamma_A \right)\left( 1 - \gamma_B\right)\displaystyle \sum_{n,m} \gamma_{A}^n\gamma_{B}^m \frac{Q_{nm}^{vu}}{4Q_{nm}},\\
p_3=p_4= &
\left( 1 - \gamma_A \right)\left( 1 - \gamma_B\right)\displaystyle \sum_{n,m} \gamma_{A}^n\gamma_{B}^m \frac{Q_{nm}^{uv}}{4Q_{nm}},
\end{eqnarray}
the sum of which is unity.

From each $\rho_i$, one obtains the mixed state of qubit B,
$\mathrm{Tr}_{A0,B0}\left(\rho_i\right)$, by tracing out the field modes.
This mixed state  is nothing but a modification of the pure state in the ideal case. After learning the measurement result of Alice on A and C,  Bob transforms the mixed state of qubit B to the corresponding destined mixed state  by using a one-qubit unitary transformation, which is the same as in the ideal case. Denote the one-qubit unitary transformation corresponding to $|\rm{Bell}^i\rangle$, which has been described above,  as $U_i$. Then the destined density matrix is
\begin{equation}
\rho^B_i = U_i \mathrm{Tr}_{A0,B0}\left(\rho_i\right) U_i^\dagger,
\end{equation}

It can be obtained that
\begin{equation}
\begin{array}{rl}
\rho^B_3=\rho^B_4= &\left( {1 - {\gamma _{A}}} \right)\left( {1 - {\gamma _{B}}} \right)\mathop \sum \limits_{n,m} \frac{{\gamma _{A}^n\gamma _{B}^m}}{{Q_{nm}^{u v }}}\{ {[ {{{| u  |}^2} + {{| v  |}^2}\left( {n + 1} \right){{| {{\mu _{A}}} |}^2} + } } \\
& {| v  |^2}\left( {m + 1} \right){| {{\mu _{B}}} |^2} +   {{{| u  |}^2}n\left( {m + 1} \right){{| {{\mu _{A}}} |}^2}{{| {{\mu _{B}}} |}^2}} ]{| 0 \rangle _{B}}\langle 0 |   \\
& + u {v ^*}{| 0 \rangle _{B}} \langle 1 | + {u ^*}v {| 1 \rangle _{B}}\langle 0 | + [ {{{| v  |}^2} + {{| u  |}^2}n{{| {{\mu _{A}}} |}^2} + } \\& { {{{| u  |}^2}m{{| {{\mu _{B}}} |}^2} + {{| v  |}^2}\left( {n + 1} \right)m{{| {{\mu _{A}}} |}^2}{{| {{\mu _{B}}} |}^2}} ]{{| 1 \rangle }_{B}}\langle 1 |} \}.
\end{array}
\end{equation}

$\rho^B_1=\rho^B_2$ is given by this expression with $u$ and $v$ exchanged.

The fidelity $F_i$ is the overlap between the destined  mixed state of qubit B  and  $|\psi\rangle$, the teleported state originally carried by qubit C, that is,
\begin{equation}
F_i=\langle \psi|\rho^B_i|\psi\rangle.
\end{equation}

It is obtained that
 \begin{equation} \begin{array}{rl}
F_3=F_4  = &\left( {1 - {\gamma _{A}}} \right)\left( {1 - {\gamma _{B}}} \right)\mathop \sum \limits_{n,m} \frac{{\gamma _{A}^n\gamma _{B}^m}}{{Q_{nm}^{u v }}}\{ {1 + {{| u  |}^2}{{| v  |}^2}\left( {2n + 1} \right){{| {{\mu _{A}}} |}^2} }  \\ &+  {\;{{| u  |}^2}{{| v  |}^2}\left( {2m + 1} \right){{| {{\mu _{B}}} |}^2} + [ {{{| u  |}^4}n\left( {m + 1} \right) + {{| v  |}^4}\left( {n + 1} \right)m} ]{{| {{\mu _{A}}} |}^2}{{| {{\mu _{B}}} |}^2}} \}. \label{f3}
\end{array}    \end{equation}

$F_1=F_2$ is given by this expression with $u$ and $v$ exchanged.

\begin{figure}[htb]
\centering
\includegraphics[width=0.8\textwidth]{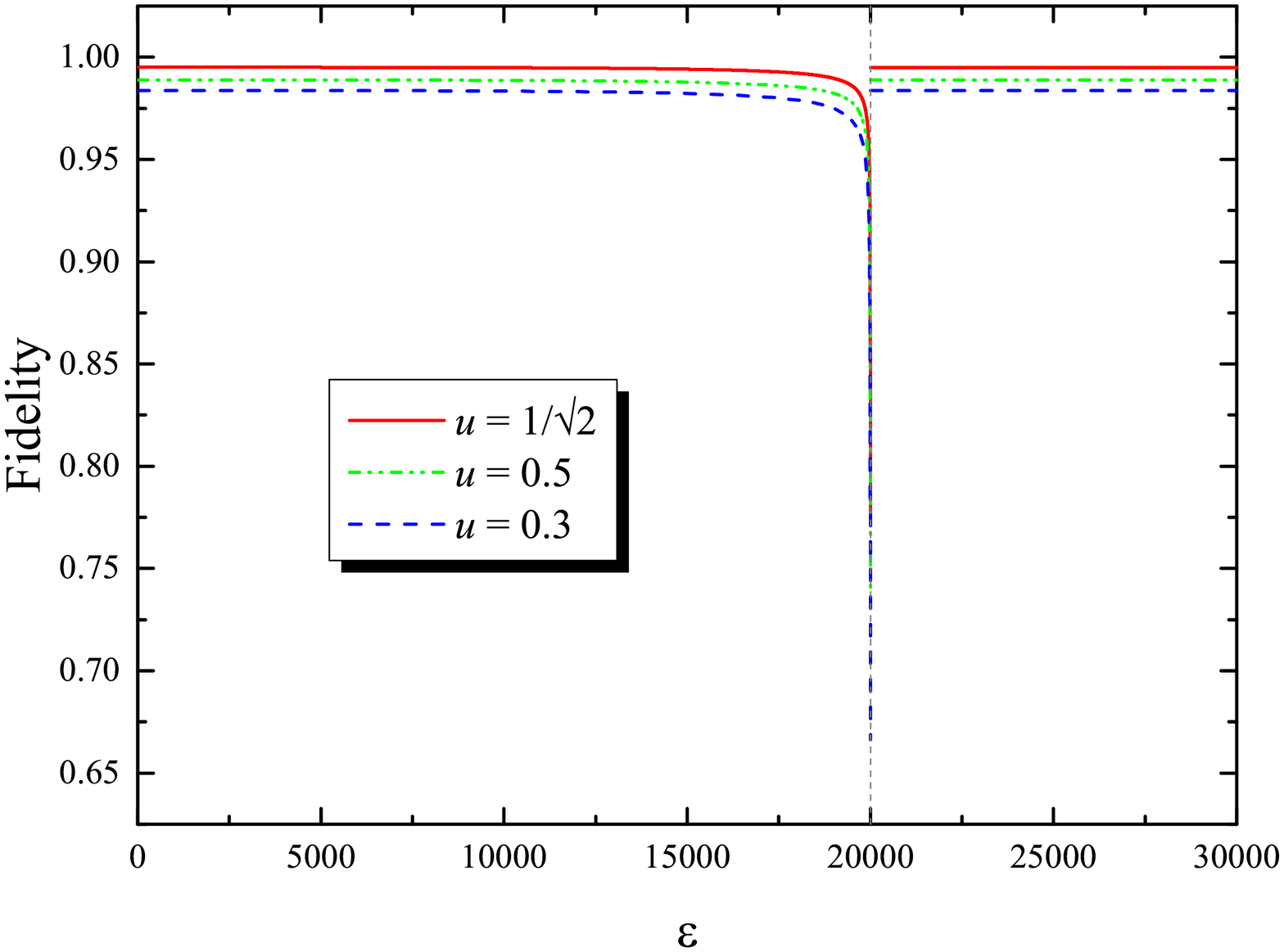}
\caption{ The fidelity of quantum teleportation, $F_3$ as defined in (\ref{f3}),  as a function of the volume  $\varepsilon$ of the expansion of the universe. The rapidity of the cosmic expansion is $\sigma=5$.  The parameters of the two originally entangled qubits A and B are the same. The energy gap of each of them is $\Omega=2$, the inner product of the mode functions, defined in Eq.~(\ref{mu0}), is $\mu  = 0.1$.  The mass of the scalar field particle is $m=0.01$.  The  state teleported from qubit C   to B is  $u|0\rangle+v|1\rangle$.
} \label{fig10}
\end{figure}

\begin{figure}[htb]
\centering
\includegraphics[width=0.8\textwidth]{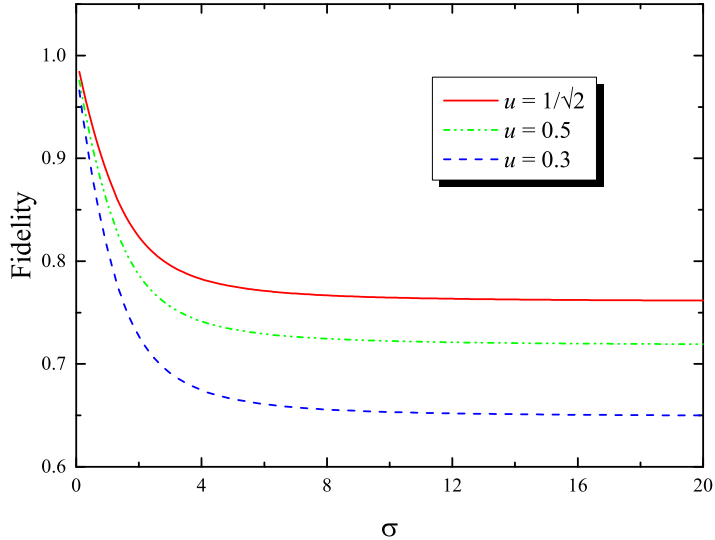}
\caption{ The fidelity of quantum teleportation, $F_3$ as defined in (\ref{f3}),  as a function of the rapidity   $\sigma$ of the expansion of the universe.     The parameters of the two originally entangled qubits A and B are the same. The energy gap of each of them is $\Omega=2$, the inner product of the mode functions, defined in Eq.~(\ref{mu0}), is $\mu  = 0.1$.  The mass of the scalar field particle is $m=0.01$. The volume of the expansion is chosen to be   ${\varepsilon  =\varepsilon _{\max }} = 19999.5$.   The  state teleported from qubit C   to B is  $u|0\rangle+v|1\rangle$.
}  \label{fig11}
\end{figure}

\begin{figure}[htb]
\centering
\includegraphics[width=0.8\textwidth]{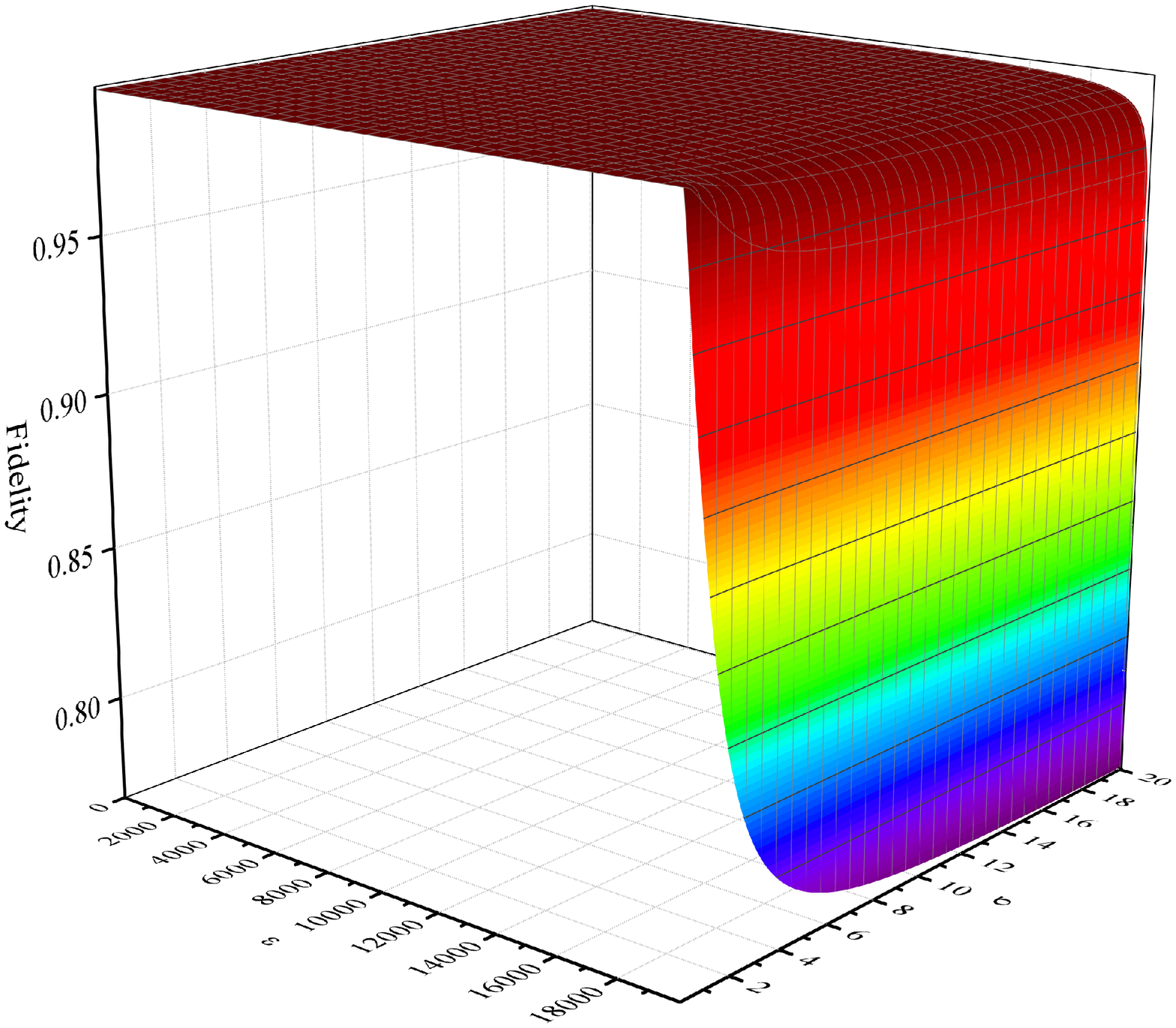}
\caption{ The fidelity of the quantum teleportation, $F_3$ as defined in (\ref{f3}),   as a function of the  volume $\varepsilon$ and  the rapidity $\sigma$ of the cosmic expansion.  The parameters of the two originally entangled qubits A and B are the same. The energy gap of each of them is $\Omega=2$, the inner product of the mode functions, defined in Eq.~(\ref{mu0}), is $\mu  = 0.1$.  The mass of the scalar field particle is $m=0.01$.  The  state teleported from qubit C   to B is  $\frac{1}{\sqrt{2}}(|0\rangle+|1\rangle)$. } \label{fig12}
\end{figure}

Only one of these four fidelities needs to be numerically calculated. It is $F_3$, for which the one-qubit unitary transformation is just unity,  that is shown in Figs.~\ref{fig10}, \ref{fig11} and \ref{fig12}. $F_4=F_3$. For the plots of $F_1$ and $F_2$, one only needs to replace $u$ as $v$ in these figures.

The dependence of the teleportation fidelity $F_3$ on the cosmic parameter   $\varepsilon$ is shown in Fig.~\ref{fig10}. For a given value of $\sigma$, the fidelity monotonically decreases with the increase of $\varepsilon$, till $\varepsilon_{max}$.

The dependence of the teleportation fidelity  $F_3$ on the cosmic  parameter $\sigma$ is shown in Fig. \ref{fig11}.   The fidelity monotonically decreases with  the increase of  $\sigma$ towards an asymptotic value.  The larger  $\sigma$, the smaller the rate of decrease of the fidelity.

For $|u| \leq 1/\sqrt{2}$ and  given values of $\varepsilon$ and  $\sigma$, the smaller $|u|$, the smaller $F_3$ and the larger $F_1$.

For $u = 1/\sqrt{2}$, the 2D plot of the teleportation fidelity  $F_3$ as a function of the two cosmic parameters $\varepsilon$ and  $\sigma$ is shown in Fig.~\ref{fig12}. In this case all the four fidelities are equal. The larger $\sigma$, the  larger the rate of decrease of  the teleportation fidelity with respect to $\varepsilon $. When $\sigma$ is small enough, the dependence of  the teleportation fidelity   on $\varepsilon$ is saturated.

In the  case that there has not expansion of the universe, the fidelity can be obtained as
 \begin{equation}
F_3=F_4  = \frac{1+ |uv|^2 ( | \mu _{A}|^2 + |\mu_B|^2) }{ 1+|v|^2( | \mu _{A}|^2 + |\mu_B|^2)},
\end{equation}
which reduces to unity when $\mu_A=\mu_B=0$. $F_1=F_2$ is given by this expression with $u$ and $v$ exchanged.

\section{Common features of different quantities \label{explain} }

One may note from the above figures that all the quantities share some similarities in their dependence on the two cosmic parameters $\varepsilon$ and  $\sigma$. In most of the range of $\varepsilon$, except near $\varepsilon_{max}$, the deviation from the static case ($\varepsilon=0$) is limited.  Certainly there are also special features. Most notable is that only concurrence can have sudden death.

The reason for the similarity is that each quantity depends on these two parameters only through $\gamma$, which is a measure of the mixture of the pair of ``in'' modes  $\mathbf{k}_0$ and $-\mathbf{k}_0$, and is a measure of the average number of particles created at the ``out''  mode  $\mathbf{k}_0$. Moreover, the following nature of the dependence of $\gamma$ on the two parameters leads to the common features in all the quantities studied above.

\begin{figure}
\centering
\scalebox{0.8}{\includegraphics{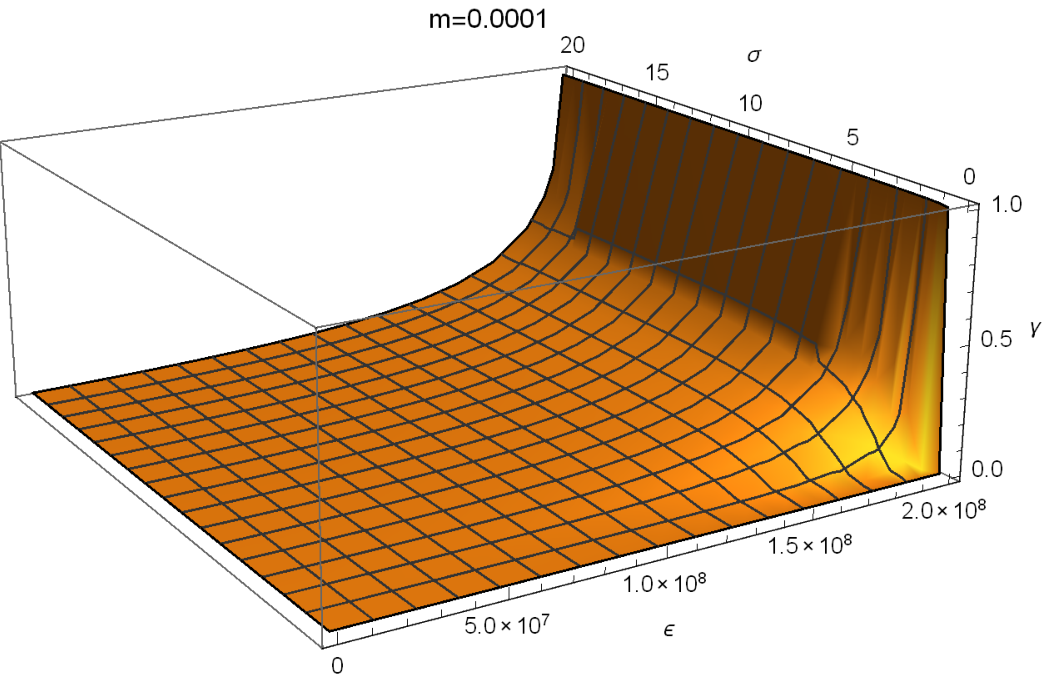}}\scalebox{0.8}{\includegraphics{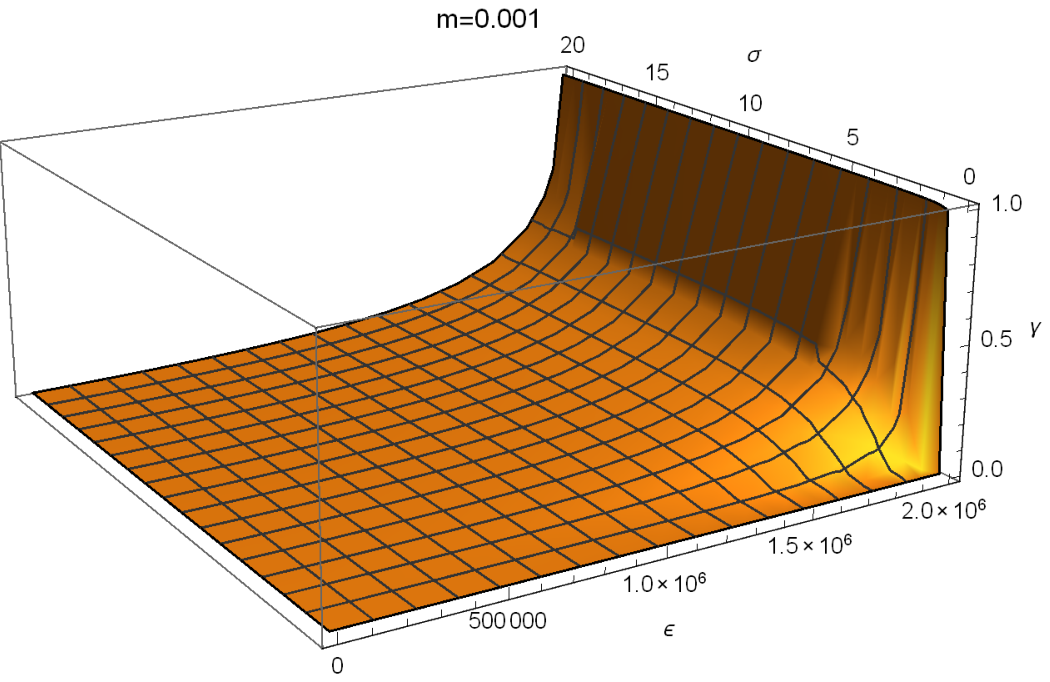}}
\scalebox{0.8}{\includegraphics{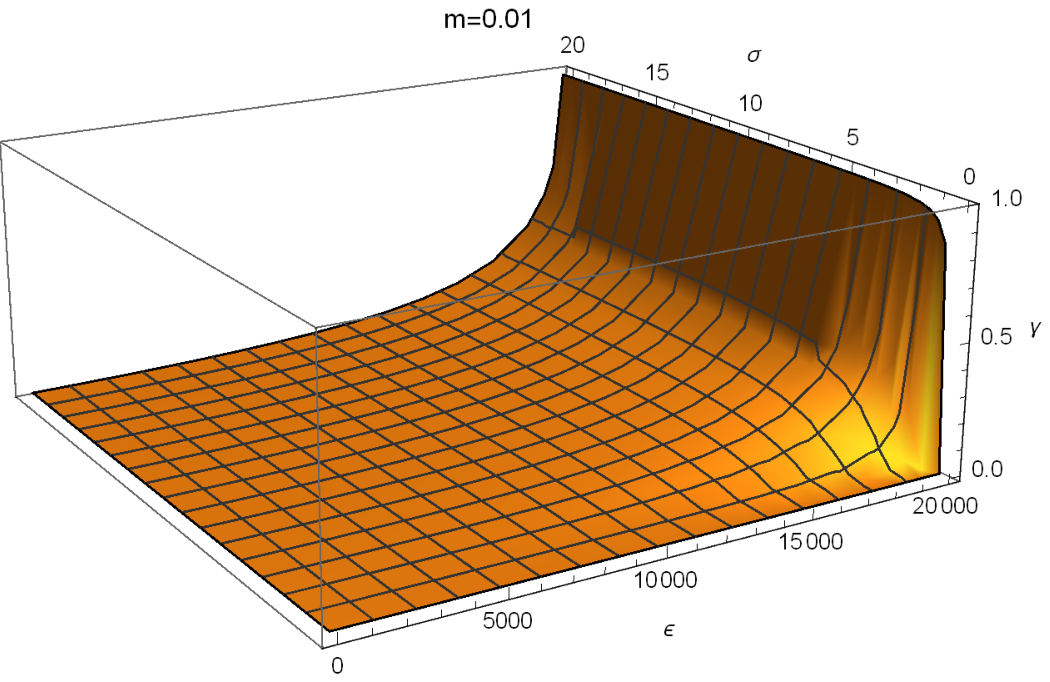}}\scalebox{0.8}{\includegraphics{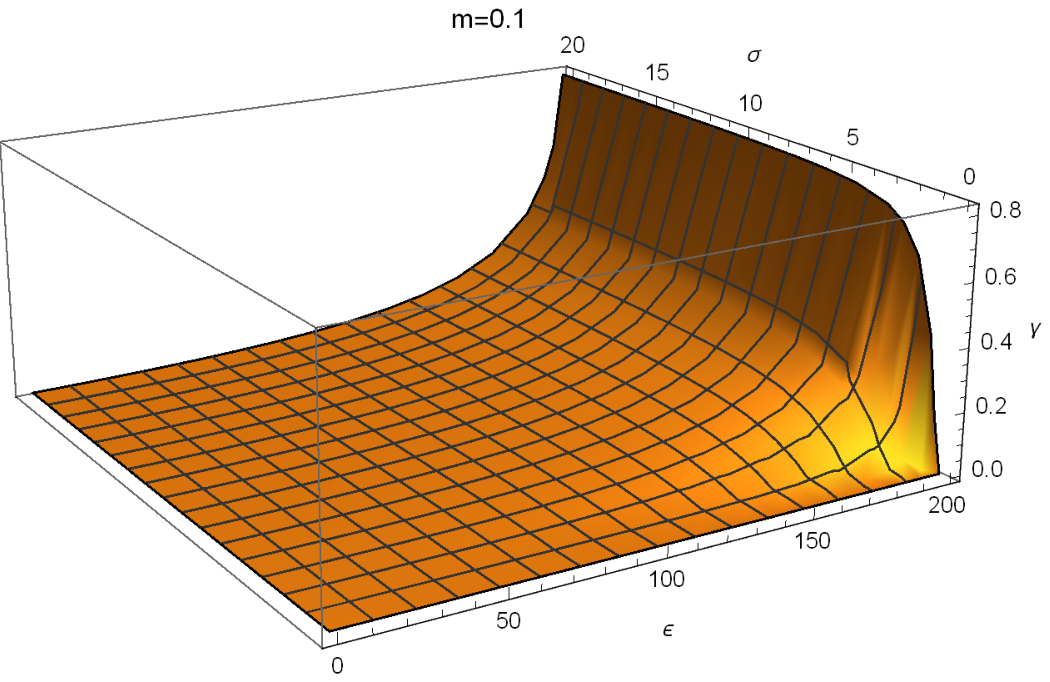}}
\scalebox{0.8}{\includegraphics{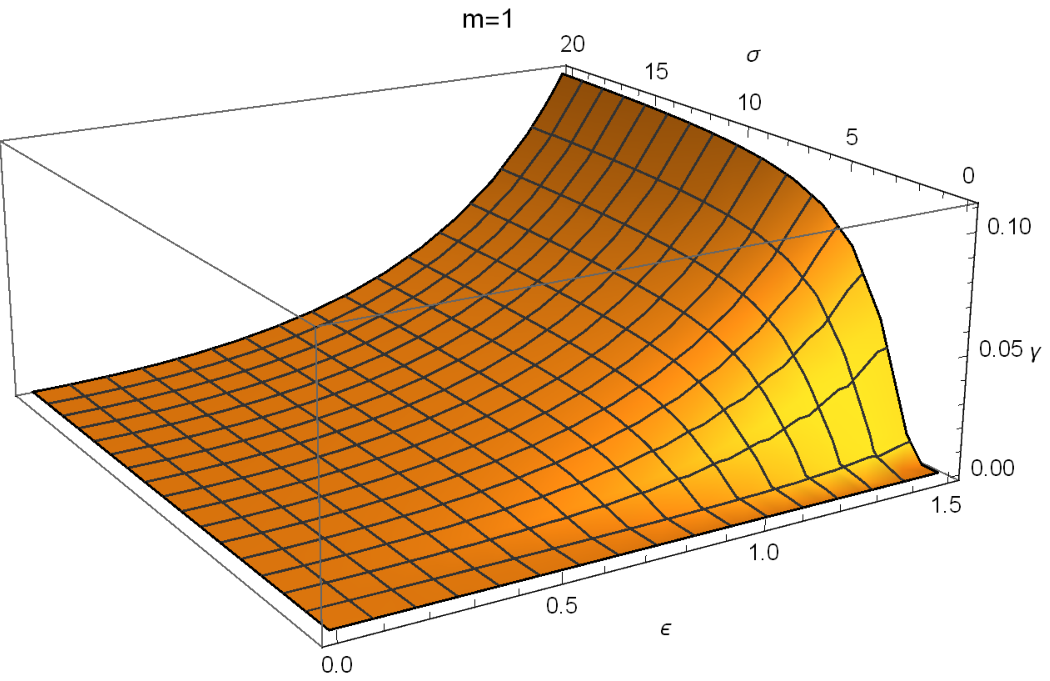}}
\caption{$\gamma$ as a function of the  volume $\varepsilon$ and  the rapidity $\sigma$ of the cosmic expansion, for $m=0.0001, 0.001, 0.01, 0.1, 1$, respectively.
The parameters of the two originally entangled qubits A and B are the same. The energy gap of the qubit is $\Omega=2$, which is supposed to be equal to the energy of the field mode $\mathbf{k}_0$ coupled with the qubit.  } \label{fig13}
\end{figure}

In Fig.~\ref{fig13}, we show $\gamma$ as a function of  $\varepsilon$ and  $\sigma$, for $m=0.0001, 0.001, 0.01, 0.1, 1$, representing five orders of magnitudes, under the constraint $\Omega \geq m$. $m=0.01$ is the value used in the above calculations. It can be seen that for each value of $m$,  $\gamma$  remains close to $0$ in a large parameter regime. The smaller $m$, the larger this regime.  Only when $m$ is of the same order of magnitude as $\Omega$, represented by $m=1$, there is a significantly large regime in which $\gamma$ is significantly larger than $0$. As depicted in Fig.~\ref{fig13}, for the other four values  of $m$, i.e. when the order of magnitude of $m$ is smaller than that of $\Omega$, $\gamma$ is significantly larger than $0$ only when $\varepsilon$ is close to $\varepsilon_{max}$. For even smaller value of  $m$, the regime in which $\gamma$ is significantly larger than $0$ is even smaller.

Consequently, all the quantities calculated above significantly deviate from the static case in a limited parameter regime.

\section{Summary and discussions \label{summary} }

In this paper, we have studied the behavior of one or two qubit detectors of scalar fields in a 1+1 dimensional conformally flat spacetime. We have compared the case that the scale factor is static with the case that the scale factor describes an expanding universe. In this model of expanding universe,  the particle content is well defined in the distant past and in the far future. It is in the far future that the detectors are considered to be coupled with the fields for a period of time.

In the static case, i.e. if there has not been expansion of the universe, the initial state of the field is assumed to be the vacuum. Unless its initial state   is the ground state  $|0\rangle$, the coupling of a single qubit with the field causes it to be decohered to a mixed state. For two initially entangled qubits A and B, their coupling with two ambient scalar fields degrades their total  correlation, quantified by the mutual information, as well as quantum entanglement, quantified by the concurrence and the teleportation fidelity. For each of these quantity,  there is a perturbative  correction  due to the qubit-field interaction, parameterized by   $\mu_A$ and $\mu_B$, as defined in Eq.~(\ref{mu0}).

In the case that there has been expansion of the universe, the vacuum sector of each pair of momenta $\pm \mathbf{k}$  in the distant past becomes in the far future  a squeezed superposition of all possible Fock states of equal occupation numbers in the pair of  modes. The particles with opposite momenta are separated on cosmological scale. A detector at the far future sees only particles with one sign, hence a mixed state of the accessible  field modes. After the interaction between the detector and the field, the state of detector becomes entangled with the field as the environment, hence the state of the detector  becomes mixed. For two entangled detectors, the interaction with the scalar fields in the expanded universe causes the degradation of the mutual information and entanglement, and even entanglement sudden death if the initial entanglement is small enough. Consequently, the fidelity of quantum teleportation based on this entanglement becomes  less than unity. Our analysis of teleportation in presence of coupling with field modes may be useful also in other areas.

We have calculated how  the purity, the mutual information, the concurrence and the teleportation fidelity depend on the two parameters characterizing  the expansion of the universe, namely the total volume $\varepsilon$ and the rapidity $\sigma$ of the expansion. It turns out that  each quantity monotonically decreases, and the rate of change increases, with the increase of  $\varepsilon$, until the maximal value $\varepsilon_{max}$, above which the field mode becomes off-resonant with the qubit. Each quantity also monotonically decreases, and the rate of change decreases, with the increase of $\sigma$. The reason for the common feature is that the dependence on  $\varepsilon$ and   $\sigma$ is only through $\gamma$, characterizing the mixing of the ``in'' modes or the number of particles created by the cosmic expansion. $\gamma$ is significantly larger than $0$  in limited parameter regimes.

Information is physical. Studying quantum informational quantities in cosmological setting  can shed new light on quantum information. With the input of quantum field theory, particle physics and gravitational physics, our understanding of quantum information will be deepened.

Physics is informational. Wheeler said: ``It from bit''. Using ideas from quantum information in cosmology can also bring new tools to the latter. For example,  these quantum informational quantities encode  information about cosmological parameters, hence it is possible to find here new probes of the universe.

One possible direction into which more realistic extension of our work can be made is inflationary cosmology, in which the exponentially expanding metric leads to the quantum fluctuation of a scalar field as the primordial inhomogeneity that has acted as the seeds for the formation of large scale structures and imprinted the cosmic microwave background. The role of the detectors in this paper could possibly be played by  some particles in post-inflation era.

Thus one may learn about the parameters of the whole universe in its history through the properties of one or a few qubits in a finite time, supplementing the traditional way of astronomical observation. This idea can be captured very fittingly by  William Blake's lines:

\begin{center}
{\it To see a World in a Grain of Sand\\And a Heaven in a Wild Flower,\\Hold Infinity in the palm of your hand\\And Eternity in an hour.}
\end{center}

\acknowledgments

This work was supported by the National Science Foundation of China (Grant No. 11374060).

\end{document}